\documentclass{aa}  
\usepackage{graphicx}
\begin{document}
\title{An investigation into the  radial velocity variations of  
  CoRoT-7 \thanks{Based on observations made with HARPS
spectrograph on the 3.6-m ESO telescope
under programs 082.C-0120, 082.C-0308(A), and DDT program 282.C-5036(A) }}

       
\author{ A.P. Hatzes\inst{1} 
\and R. Dvorak\inst{2}
\and G. Wuchterl \inst{1}
\and P. Guterman\inst{3}
\and M. Hartmann \inst{1}
\and M. Fridlund \inst{4}
\and D.~Gandolfi \inst{1,}\inst{4}
\and E.~Guenther \inst{1}
\and M.~P\"atzold \inst{5}
}

\institute{ 
Th\"uringer Landessternwarte Tautenburg, Sternwarte 5, 07778 Tautenburg, Germany
\and
Institute for Astronomy, University of Vienna, T\"urkenschanzstrasse 17, 1180 Vienna, Austria
\and
Laboratoire d'Astrophysique de Marseille, UMR 6110, Technopole de
Marseille-Etoile, F-13388 Marseille cedex 13, France
\and
Research and Scientific Support Department, European Space Agency, ESTEC, 2200 Noordwijk, The Netherlands
\and
Rheinisches Institut f\"ur Umweltforschung, Universit\"at zu K\"oln, Abt. Planetenforschung, Aachener Str. 209, 50931 K\"oln, Germany
}

\offprints{
  Artie Hatzes, \email{artie@tls-tautenburg.de}\\
  \\}

   \date{Received; accepted}

 
  \abstract
{
CoRoT-7b, the first transiting ``superearth''  exoplanet, has a radius of
1.7 $R_{\oplus}$ and a mass of 4.8 $M_{\oplus}$. The HARPS radial velocity (RV)
measurements used for deriving this mass also detected
an  additional companion with a period of 3.7 days
and a mass of 8.4 $M_{\oplus}$. The mass of CoRoT-7b is a crucial
parameter for planet structure models, but is difficult to determine because
CoRoT-7 is a modestly active star and there is at least  one
additional companion.} 
{The aims of this paper are to assess the statistical significance of the
RV variations of CoRoT-7b and CoRoT-7c, to obtain  a better
measurement of the planet mass for CoRoT-7b, and to search for additional
companions in the RV data.}
{A Fourier analysis is performed on the HARPS spectral data
of CoRoT-7. These data include RV measurements,
spectral line bisectors, the full width at half maximum of the cross-correlation
function, and Ca II emission. The latter 3 quantities vary due to stellar
activity and were used to assess the nature of the observed RV
variations. An analysis of a sub-set of the RV measurements where
multiple observations were made per night was also used to estimate
the RV amplitude from CoRoT-7b that was less sensitive to activity
variations.}
{Our analysis indicates that the 0.85-d and 3.7-d RV signals
 of CoRoT-7b  and CoRoT-7c are present in the spectral data
with a high degree of statistical significance. We also find evidence for 
another significant 
RV signal at 9 days. An analysis of the activity indicator
data reveals that this 9-d signal  most likely does not arise from activity,
but possibly from an additional companion. If due to a planetary
companion the mass is m $=$ 19.5 $M_{\oplus}$, assuming co-planarity with
CoRoT-7b. 
A dynamical study of the three planet system shows that it is stable over 
several hundred millions of years.
 Our analysis yields a RV amplitude of 5.04 $\pm$ 1.09 m\,s$^{-1}$  for
 CoRoT-7b which 
corresponds to a planet mass of m = 6.9 $\pm$ 1.4 $M_{\oplus}$. This increased
mass 
would make the planet CoRoT-7b more Earth-like in its internal structure.
}
{CoRoT-7 is confirmed to be a  planet  system with at least 2 and possibly
3 exoplanets having masses in the range 7--20 $M_{\oplus}$.  If the third
companion can be confirmed then CoRoT-7 may represent a case of an
ultra-compact planetary system. 
}


\titlerunning{An Investigation of CoRoT-7}
\maketitle

\section{Introduction}

\begin{figure*}
\resizebox{\hsize}{!}{\includegraphics{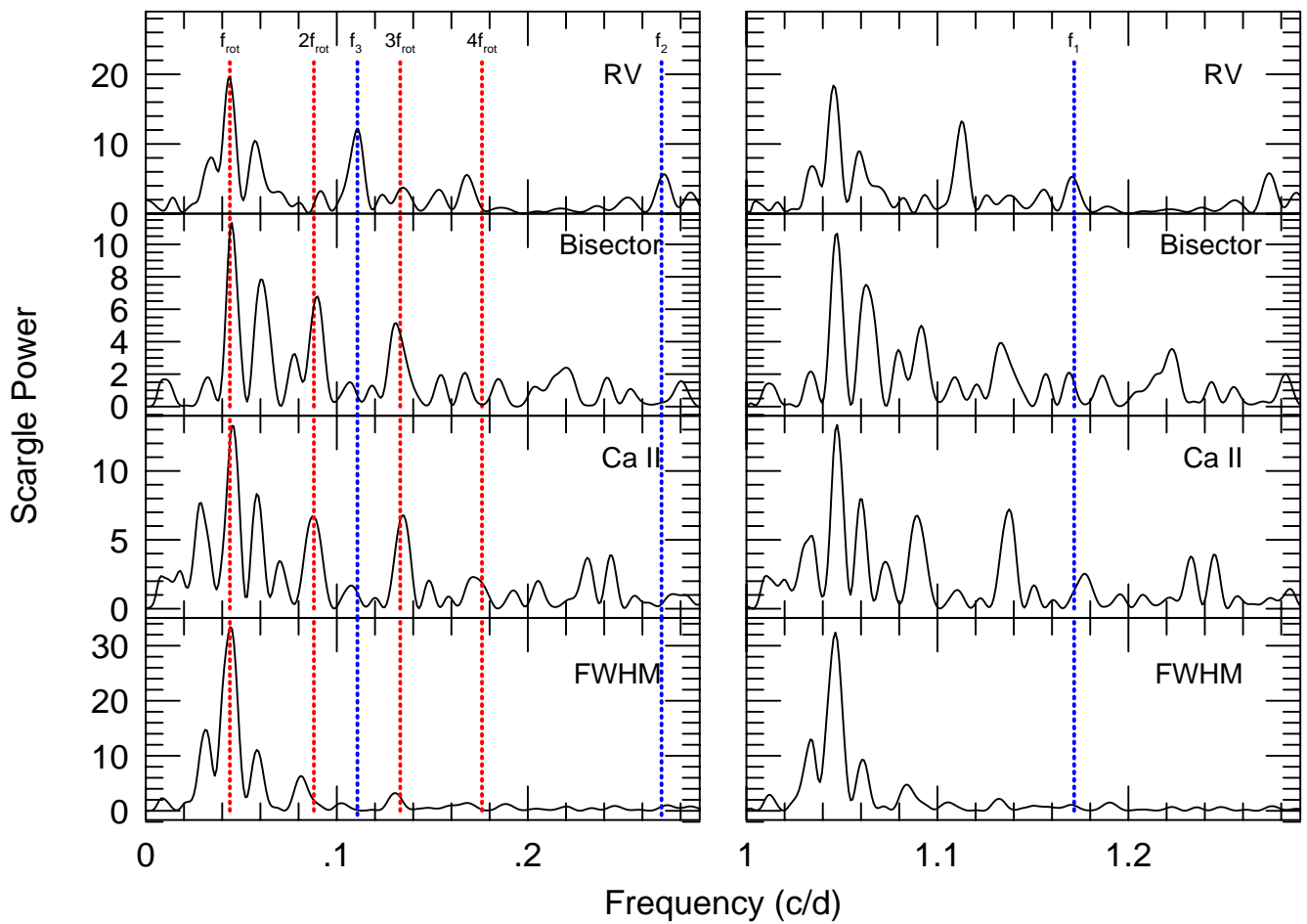}}
\caption{Scargle periodograms of the RV (top), bisector (upper middle), 
Ca II S-index (lower middle), and FWHM (bottom) measurements for the frequency range 0--0.3 c\,d$^{-1}$ 
(left panels) and 1.0--1.3 c\,d$^{-1}$ (right panels). The vertical blue lines
mark frequencies seen only in the RV (the one in the right panel is the
transit frequency). Vertical red  lines mark the rotational frequency and its
first 3 harmonics.
}
\label{scargles}
\end{figure*}

The 27-cm space telescope CoRoT is devoted to obtaining ultra-high 
precision light
curves over a wide field for the dual purpose of asteroseismology and the
detection of transiting exoplanets (Baglin 2006; Auvergne et al. 2009).
This mission resulted in the milestone discovery of 
CoRoT-7b,  the first  transiting superearth whose radius and mass
have been accurately characterized. Photometric measurements of CoRoT-7
made with the 
CoRoT Space Telescope from October 2007 to March 2008 revealed a transit like event
with a period of 0.85 days and a depth of a mere 0.03\%. A detailed analysis
of the CoRoT light curve and ancillary  measurements provided by ground
based observations excluded all sources of false positives and established
with high probability that the transit event was due to a planet with a
radius of 1.7 $R_{\oplus}$ (Leger et al. 2009). 

 CoRoT-7 is a G9 main sequence star with an estimated age of
1.2--2.3 Gyr. 
From the analysis of the three best spectra obtained with HARPS, and using 
several methods, Bruntt et al. (2010) find $T_{\rm eff}$ = 5250 $\pm$ 60 K, log g = 
4.47 $\pm$ 0.05, [M/H] = +0.12 $\pm$ 0.06, and $v$ sin $i$ = 1.1 $\pm$ 1.0 
km\,s$^{-1}$. 
They also find slightly different values for the mass and radius of CoRoT-7: 
0.91 $\pm$ 0.03 $M_\odot$  and 0.82 $\pm$ 0.04 R$_\odot$. 
The revised stellar radius results in a 
slightly smaller radius for the 
planet  of 1.58 $\pm$ 0.10 $R_{\oplus}$.

Queloz et al. (2009; hereafter Q09)  presented over 100 precise stellar radial velocity (RV) measurements
of CoRoT-7 taken between November 2008 and February 2009 with the 
High Accuracy Radial velocity Planet Searcher
(HARPS) spectrograph mounted on ESO's 3.6m telescope
at La Silla. The RV analysis presented in Q09 was complicated by the relatively
high activity of the host star. The CoRoT-7 light curve shows photometric variations of
up to 2\% with a rotation period of 23 days. The implied spot filling factor
suggests an RV ``jitter'' due to stellar activity of more than 10 m\,s$^{-1}$
(Saar \& Donahue 1997) which was confirmed by the RV measurements.

Two approaches were used in Q09 for the analysis
of the RV data. 
Fourier pre-whitening resulted in an RV amplitude of 4.16 $\pm$ 0.27 m\,s$^{-1}$
for CoRoT-7b, while filtering using rotational
harmonics resulted in an amplitude of  1.90 $\pm$ 0.4 m\,s$^{-1}$. By applying correction
terms to these amplitudes due to the effects of the filtering process  
 a consistent amplitude of 3.5 $\pm$ 0.6 m\,s$^{-1}$ was obtained.
This corresponded to a planet mass of 
 4.8 $\pm$ 0.8 $M_{\oplus}$.  Their analysis
also  revealed the presence of a 3.7-d period with an amplitude
of  4.0 $\pm$ 0.5 m\,s$^{-1}$ that was not associated with
the stellar activity. This signal is most likely due to a second companion with a mass
of 8.4 $\pm$ 0.9 $M_{\oplus}$.

One of the most important parameters we can determine for an exoplanet is its mass.
For transiting exoplanets we also know the radius and the 
average density of the planet and this is a unique parameter that  will allow
us to estimate the planet's composition, e.g. the fraction of metals and/or
water. For transiting exoplanets we also know the radius and average density of
the planet, $\rho_{\rm avg}$.
A  high value of $\rho_{\rm avg}$, as seem to be the case for CoRoT-7b (Q09, this
work), indicates a high metal content and a low water abundance.
Interestingly,  another low-mass planet with a similar radius
(2.7 $R_{\oplus}$) has recently been found by the MEarth project (Charbonneau et al. 2010), but with a
 much lower $\rho_{\rm avg}$ than CoRoT-7b. Valencia et al. (2010) using the 
 mass of CoRoT-7b from Q09
concluded that the internal structure was consistent with
 a significantly depleted iron core. However, a mere increase of  only
 1-$\sigma$ in the mass and a 
 slightly smaller radius would make the planet more Earth-like. 
Clearly, a careful determination of the mass, to go with the radius
provided by the exquisite CoRoT photometric data is absolutely imperative. 
This mass must rely on how accurately we can determine the RV amplitude of CoRoT-7b
and unfortunately the activity signal makes this difficult. The details in which
the activity signal is removed may affect the final RV amplitude of CoRoT-7b.

The HARPS RV data for CoRoT-7 show complex variations due to 
multi-periodic signals. Disentangling RV signals due to companions
from that due to  activity is challenging. This is particularly true because
activity in the form of spot evolution, migration, etc. coupled with our 
poor temporal sampling
can introduce Fourier frequencies other than rotation frequency.
In this paper we focus on a more extensive frequency analysis of the RV data
that could only partially be presented in Q09. We also include in our analysis
ancillary data on the activity indicators for this star. The main
goals of this follow-up paper are:

\begin{itemize}

\item To assess the statistical significance of CoRoT-7b seen in the
RV data

\item To look for possible additional planet signals in the HARPS data

\item To understand the nature of all detected signals in the HARPS 
RV data

\item To obtain the best possible determination of the mass, and thus density of CoRoT-7b.

\end{itemize}

\section{Observations}

The HARPS spectrograph obtained 106 RV measurements for
CoRoT-7 over a time
span of 4 months. We refer the reader to Q09 for a detailed description of the
RV measurements. 
Besides RV information, the HARPS data also provided information
on the activity of the star via the bisector span of the cross
correlation function (CCF), the Ca II S-index, and the 
full-width at half maximum
(FWHM) of the CCF. In particular, the line  bisector
has become a common tool for confirming exoplanet discoveries
(e.g. Hatzes, Cochran, \& Bakker 1988; Queloz et al. 2001).
Fig. 1 of Q09 shows the RV measurements for CoRoT-7 as well as the activity
indicators.

\section{Scargle periodograms of measured quantities}

A periodogram analysis can give us a quick overview  as to possible periods
that may be present in the data.
Scargle periodograms were calculated for the four  quantities measured
from the HARPS spectra: RV, bisector span, 
Ca II S-index, and the FWHM of the  CCF.
These are shown in Figure~\ref{scargles}. The RV measurements 
contain information about possible planetary companions {\it and} activity
while the other quantities should only contain information 
on stellar activity. A comparison of the periodograms gives a first indication
as to which peaks in the RV periodogram arise from
activity and those which may be due to companions.
Two frequency ranges are shown. The left panels are for 0 $<$
$\nu$ $<$  0.3 c\,d$^{-1}$  and the right panels are for 
1.0 $<$ $\nu$  $<$ 1.3 c\,d$^{-1}$.  The vertical blue lines indicate RV
frequencies not seen in the activity indicators. The one in the right panel
is the CoRoT-7b transit frequency.  Because of 1-day aliases all frequencies
in the left panel also appear as peaks at $\nu$ $+$ 1 in the right panel.
(Note: we have not marked the 1-day alias of the CoRoT transit frequency
at 0.17 c\,d$^{-1}$, although one can see a peak at this frequency location).
The vertical red  lines mark the
rotational frequency, $f_{\rm rot}$, and its first 3 harmonics (2$f_{\rm rot}$,
3$f_{\rm rot}$, and 4$f_{\rm rot}$).

There are several important features to note in these periodograms. First,
the dominant peak in all 3 quantities occurs at $\nu$ = 0.044  c\,d$^{-1}$,
the rotational frequency as determined from the CoRoT light curve. 
Clearly, RV variations are dominated by the activity
RV jitter from rotational modulation which will complicate the extraction
of RV variations due to bona fide companions.
Second, the periodograms of the bisectors and
Ca II look remarkably similar with the same peaks identifiable in both periodograms.
Third, although the FWHM periodogram shows similar peaks near the
rotational frequency, its shape looks more like the RV periodogram
{\it but without the peaks shown by the vertical dashed lines.} The
strong peaks at $\nu$ = 0.09 and 0.13 c\,d$^{-1}$, which are the first
and second harmonics of $f_{\rm rot}$, are not as strong in the FWHM as in the
bisector and Ca II periodogram.
The most important point is that the 
3 peaks seen in the RV data at $\nu$ = 0.11, 0.27, and 1.17
c\,d$^{-1}$ (and their 1-day aliases) are not found in any of the other
quantities. This is our first hint for the presence of RV variability not
associated with 
stellar activity.

\begin{figure}
\resizebox{\hsize}{!}{\includegraphics{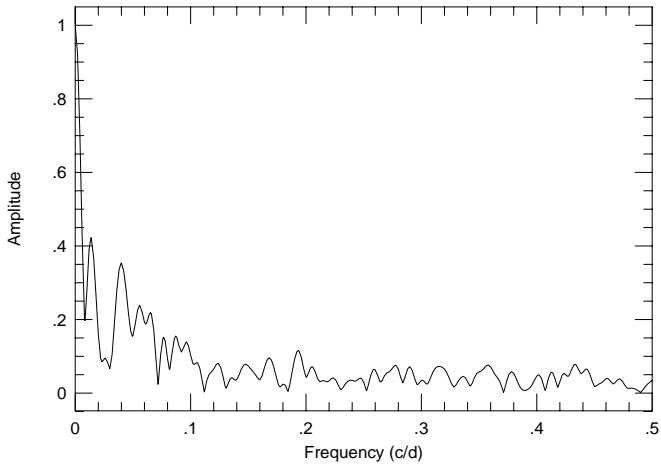}}
\caption{Spectral window function for the HARPS measurements.
}
\label{specwin}
\end{figure}

Figure~\ref{specwin} shows the spectral window for the data. As expected it
is rather complex with strong sidelobes at +0.013 and 0.04 c\,d$^{-1}$. 
Note that the same spectral window applies for all data (RV, bisectors,
Ca II, and FWHM).

\section{Frequency analysis of the data sets}

It is evident from the periodograms that all spectral quantities have
multi-periodic variations. As such they are well suited to the classic
technique of pre-whitening often employed in the study of stellar oscillations.
In this process a  Fourier transform (FT) finds the
highest peak in the power spectrum. 
A least squares fit to the frequency, amplitude, and phase to the found
signal  is made and then subtracted from the time series.
 Note that by subtracting this signal we also remove all 
aliases of the dominant frequency. A subsequent FT on the residuals
yields the next dominant frequency in the time series. This sequential
subtraction of dominant components is  continued until the level of the
noise is obtained using the criterion that peaks more than 4 times
the Fourier noise level are regarded as real (Kuschnig et al. 1997).
This frequency analysis was performed on the HARPS
data using the program {\it Period04 } (Lenz \& Breger 2004).

\subsection{Radial velocity data}

A pre-whitening analysis of the RV measurements was already
presented in Q09. In that analysis a 12-component sine
fit was made to the data which resulted in an rms to the
fit of 1.81 m\,s$^{-1}$. 
Table 1 lists frequencies using a more conservative solution. In this case
the stopping criterion for the pre-whitening (last significant peak)
was assessed using a bootstrap randomization procedure (e.g. K\"urster
et al. 1997). Frequencies above the horizontal line have a false
alarm probability less than or equal to about 1\%. The rms scatter
of the fit to the RV  is 2.96 m\,s${^-1}$ which is significantly higher
than the mean RV error of 1.89 m\,s$^{-1}$. Including the two frequencies 
below the horizontal line results in an rms scatter of 2.18 m\,s$^{-1}$,
only slightly worse than the 12-component fit from Q09. Fig. 5 in Q09
shows the fit to the RV data provided by the 12-component fit. The 9-component
fit listed in Table 1 is of comparable quality so is not shown here.

Most of the frequencies from the FT analysis arise from stellar activity
with the most dominant one being the rotational frequency, $f_{\rm rot}$ = 
0.0448 c\,d$^{-1}$ and its harmonics. This is noted in the comment column.
The rotational period derived from the RV data is 22.32 $\pm$ 0.08 days, 
hence larger than the value of 21.65 $\pm$ 0.03 days obtained from the CoRoT
light curve. 
It is not known whether this  
difference merely reflects the difficulty in measuring the rotational period
from such complex RV and light curves, or if this is a real difference
due to possibly differential rotation.
There are 3 frequencies denoted $f_1$, $f_2$, and $f_3$ that are not readily
associated with stellar activity. The frequency $f_1$ is the CoRoT-7b transit
frequency of 1.17165  c\,d$^{-1}$, $f_2$ is the 3.7-day planet reported
by Q09, and $f_3$ corresponds to a period of 9.02 days and will discussed at
length below. 
In Table 1 there are three additional frequencies above
the horizontal line and two below that could not be readily associated
with rotational harmonics, but as we show below these are most likely
related to the activity signal. For convenience in referencing we denote
sequentially the additional frequencies $f_4$ - $f_8$, although
for the following discussion $f_1$, $f_2$, and $f_3$ are the most important.

\begin{table}
\begin{center}
\begin{tabular}{cccc}
\hline
Frequency   &  Amplitude  & Comment  \\
(c\,d$^{-1}$)       &     (m\,s$^{-1}$)&  \\
\hline
\hline
0.0448  $\pm$ 0.00016 & 8.68 $\pm$  0.46    &  $f_{\rm rot}$ \\
0.1108	$\pm$ 0.00021 & 6.94 $\pm$ 0.45    &  $f_3$ \\
0.0959	$\pm$ 0.00025 & 5.92 $\pm$ 0.43    &  $f_4$ \\
1.1702	$\pm$ 0.00026 & 5.75 $\pm$ 0.44    &  $f_1$\\
0.2707	$\pm$ 0.00029 & 5.10 $\pm$ 0.40    &  $f_2$ \\
0.0353	$\pm$ 0.00028 & 4.34 $\pm$ 0.49     & $f_5$ \\
0.1815	$\pm$ 0.00046 & 3.57 $\pm$  0.44     & $f_6$ \\
\hline
0.2516	$\pm$ 0.00080 & 2.23 $\pm$ 0.40     & $f_7$ \\
0.0868  $\pm$ 0.00080 & 2.03 $\pm$ 0.40     & $f_8$ \\
\hline
\end{tabular}
\caption{Frequencies and amplitudes found by a pre-whitening procedure
on the HARPS RV data. Frequencies above the horizontal line have false alarm
probabilities less than 0.01. Those below the horizontal line are additional
frequencies found by continuing the pre-whitening procedure that resulted
in an rms scatter close to the mean measurement error of the HARPS RV
data.
}
\end{center}
\end{table}

The CoRoT light curve for CoRoT-7 is complex and shows
evidence for spot evolution on time scales of less than
150 days, or comparable to the time span of our RV
measurements. Although the fit to the data using
the full RV data set is excellent, the Fourier sine components may not
accurately represent the activity variations over such 
a long time span. An analysis of the data in subsets of shorter time interval may minimize
the uncertainties in the activity signal due to spot evolution.

To minimize possible effects of spot evolution, a frequency analysis was also performed on sub-sections of the
HARPS data. The RV data were divided into  3 sets each roughly spanning
one rotation period. The time span of these data sets
are listed in the first 3 lines of Table 2.  When
analyzing data containing periods comparable to the 
 time span of the data string there can be risks to  using the pre-whitening procedure.
 This is certainly true for the rotational period and to some extent to the
9-d period since it only undergoes 2 cycles over the time span of the subset observations.
Likewise the 0.85-d suffers from being severely under-sampled for some of 
the data subsets. In short, 
alias effects and spectral leakage may result in a spurious period being identified
and removed from the data and thus resulting in an erroneous solution. 

Indeed, when performing a straight  pre-whitening procedure to Subset 1 the dominant
frequency is $f$ = 0.13 c\,d$^{-1}$ and the rotational frequency appears
at 0.05 c\,d$^{-1}$. Only $f_2$ is recovered. However, when one first
fits the data using the CoRoT-7b frequency first, the procedure recovers
both $f_2$ and $f_3$. A pre-whitening analysis on Subset 2 recovers $f_{\rm rot}$,
$f_1$, and $f_2$, but the nearest frequency to $f_3$ was at 0.08 c\,d$^{-1}$.
Subset 3 also yields different results. The dominant frequency occurs at 0.03 c\,d$^{-1}$,
significantly different from $f_{\rm rot}$. Both $f_2$ and $f_3$ are recovered, but the nearest frequency
to $f_1$, the transit frequency, is at 1.18 c\,d$^{-1}$.

Therefore, in fitting the subset
data we assumed   that $f_1$, $f_2$, and $f_3$ were present in the data
 asked the question ``Can the subset
data be fit using  these frequencies?" To answer this we first
fit and removed the frequencies that were found in the full data set,
namely  $f_{\rm rot}$, $f_1$, $f_2$, and $f_3$. In this fitting only the frequencies were
held fixed, but the amplitudes and phases were allowed to vary. The pre-whitening
procedure was continued until the rms fit to the data was comparable to the HARPS
measurement error (better than 2 m\,s$^{-1}$). 

Table 3--5 lists the frequencies and amplitudes
from the Fourier analysis. In all cases the rms fit of the data is comparable
to the mean error of the HARPS measurements.
The amplitude of the
$\nu$ = 0.11, 0.27, and 1.17 c\,d$^{-1}$  signals remain relatively constant.
The largest amplitude variations are found for the frequency associated
with rotation, but it is not clear how significant these are given the
data sub-interval spans only one rotation period. The amplitude is thus
 probably not well determined. In summary, this has established
that the RV variations in the subset data are consistent with the
presence of $f_1$, $f_2$, and $f_3$, and with amplitudes comparable to those
found in the full data set. Note that the additional frequencies found in the
pre-whitening procedure can be identified with frequencies found in the full
data set analysis. 

The referee suggested an analysis of the longest data string between
JD = 2454847.6 - 2454884.7 as this has few gaps (Subset 4). A Fourier pre-whitening
procedure found $f_1$ and  $f_2$, but was unable to find $f_3$. This was also
the case when 
we generated  a fake data set using the periods listed in Table 1, sampled
in the same manner as Subset 4, and  with random noise with $\sigma$ = 2
m\,s$^{-1}$. The pre-whitening procedure was also unable to detect $f_3$
even though it was present in the data. Table 6 lists the results of 
the Fourier analysis by first fitting $f_{\rm rot}$, $f_1$, $f_2$, and $f_3$
- similar to the procedure used for generating Tables 3-5. Note that the
amplitude of  $f_3$ is considerably lower than was found for Subsets 1-3.
This may be an indication that it is an artifact of the activity signal.
However, when performing the same analysis on the fake data generated
using the frequencies and amplitudes in Table 1 (again with
the appropriate sampling and noise), the fitted amplitude for
$f_3$ is a factor of 3 less.   The fitted amplitudes for the fake data
are listed in the third column of Table 6.
So, any evidence for amplitude variations
for $f_3$ is inconclusive.

\begin{table}
\begin{center}
\begin{tabular}{ccc}
\hline
Data   & Range (JD) \\
\hline
\hline
Setset 1  & 2454775.8 - 2454807.8 \\
Setset 2  & 2454825.7 - 2454860.7 \\
Setset 3  & 2454861.7 - 2454884.7 \\
Setset 4  & 2454847.6 - 2454884.7 \\
\hline
\end{tabular}
\caption{The data subsets.}
\end{center}
\end{table}

\begin{table}
\begin{center}
\begin{tabular}{ccc}
\hline
Frequency   &  Amplitude  & Comment \\
(c\,d$^{-1}$)  &  (m\,s$^{-1}$)&   \\
\hline
\hline
0.0447   &  9.23  $\pm$ 0.55 & $f_{\rm rot}$ \\
0.1108   &  4.97  $\pm$ 0.52 & $f_{3}$\\
0.2707   &  5.41  $\pm$ 0.48 & $f_{2}$\\
1.1716   &  4.74 $\pm$ 0.52  & $f_{1}$\\
0.0218   &  8.97   $\pm$ 0.57&                       \\
0.2203   &  1.92  $\pm$ 0.58 &                      \\ 
\hline
$\sigma$ = 1.98 m\,s$^{-1}$ & \\
\hline
\end{tabular}
\caption{Frequencies found in data Subset 1}
\end{center}
\end{table}

\begin{table}
\begin{center}
\begin{tabular}{ccc}
\hline
Frequency   &  Amplitude & Comment   \\
(c\,d$^{-1}$)        &    (m\,s$^{-1}$) & \\
\hline
\hline
0.0447   &  5.23 $\pm$ 0.41 & $f_{\rm rot}$ \\
0.1108   &  7.75 $\pm$ 0.40 & $f_3$  \\
0.2707   &  5.60 $\pm$ 0.34 & $f_2$ \\
1.1716   &  5.25 $\pm$ 0.38 & $f_1$  \\
0.0227   &  6.42 $\pm$ 0.38 & \\
0.1342   &  7.94 $\pm$ 1.35 & \\
\hline
$\sigma$ = 1.89 m\,s$^{-1}$ & \\
\hline
\end{tabular}
\caption{Frequencies found in data Subset 2}
\end{center}
\end{table}

\begin{table}
\begin{center}
\begin{tabular}{ccc}
\hline
Frequency   &  Amplitude   \\
(c\,d$^{-1}$)    &  (m\,s$^{-1}$)    & \\
\hline
\hline
0.0447   & 12.80 $\pm$ 0.61 & $f_{\rm rot}$  \\
0.1108   &  5.08 $\pm$ 0.57  & $f_{3}$  \\
0.2707   &  5.87 $\pm$ 0.44 & $f_{2}$   \\
1.1716   &  6.10 $\pm$ 0.41 & $f_{1}$    \\
0.080    &  4.02 $\pm$ 0.70&  \\
\hline
$\sigma$ = 1.99 m\,s$^{-1}$ & \\
\hline
\end{tabular}
\caption{Frequencies found in data Subset 3}
\end{center}
\end{table}

\begin{table}
\begin{center}
\begin{tabular}{ccc}
\hline
Frequency   &  Amplitude  &  Amplitude (fake)  \\
(c\,d$^{-1}$)    &  (m\,s$^{-1}$)  & (m\,s$^{-1}$)    \\
\hline
\hline
0.0447   & 13.79 $\pm$ 0.45 & 13.30  $\pm$ 0.43  \\
0.1108   &  2.09 $\pm$ 0.47 & 2.75   $\pm$ 0.46  \\
0.2707   &  5.90 $\pm$ 0.35 & 6.33   $\pm$ 0.42  \\
1.1716   &  4.54 $\pm$ 0.40 & 4.29   $\pm$ 0.35    \\
0.085    &  4.75 $\pm$ 0.52 & 5.21   $\pm$ 0.42  \\
\hline
$\sigma$ = 2.34 m\,s$^{-1}$ & \\
\hline
\end{tabular}
\caption{Frequencies found in data Subset 4 using the same
analysis as Tables 3-5,  i.e. first removing the contribution of
$f_rot$, $f_1$, $f_2$, $f_3$. Column 3 shows the amplitude
derived using fake data (see text).}
\end{center}
\end{table}

\subsection{Frequency analysis of activity indicators}

A Fourier analysis with pre-whitening
was performed on the bisector,  Ca II S-index, and FWHM
measurements. These are listed in Tables 7--9. 
The pre-whitening procedure was continued beyond the last frequency
we considered significant 
to see if at any point  a frequency was detected that coincided
with either $f_1$, $f_2$, and $f_3$.
Both the bisector and FWHM measurements show frequencies near
$f_2$: $\nu$ = 0.28 c\,d$^{-1}$  ($P$ = 3.6 d), but only after 
pre-whitening the data well past our stopping criterion
(7th frequency found in the bisector and the 6th frequency
in the FWHM). These amplitude frequencies are essentially at or below the noise level.
Furthermore, when
we phase the bisector and FWHM data after removing all components except those
near the 3.7-d period, no obvious sinusoidal variations are present. We therefore
do not consider these frequencies to be significant. 
\begin{table}
\begin{center}
\begin{tabular}{ccc}
\hline
Frequency   &  Amplitude &   Comment  \\
(c\,d$^{-1}$)        &       (m\,s$^{-1}$)&  \\
\hline
\hline
0.0432   &  4.21 & $f_{\rm rot}$ \\
0.0905   &  2.81 & $f_4$   \\
0.0667   &  3.00 &  \\
\hline
\end{tabular}
\caption{Frequencies found in the bisector data using a pre-whitening 
procedure. }
\end{center}
\end{table}

\begin{table}
\begin{center}
\begin{tabular}{ccc}
\hline
Frequency   &  Amplitude &   Comment  \\
(c\,d$^{-1}$)        &                    &   \\
\hline
\hline
0.0447  &  0.0159  & $f_{\rm rot}$ \\
0.0128  &  0.0088  &     \\
0.1369  &  0.0089  &  \\
\hline
\end{tabular}
\caption{
Frequencies found in the Ca II data using a pre-whitening
procedure. }
\end{center}
\end{table}

\begin{table}
\begin{center}
\begin{tabular}{ccc}
\hline
Frequency   &  Amplitude &   Comment  \\
(c\,d$^{-1}$)        &       (km\,s$^{-1}$)&  \\
\hline
\hline
0.044  &  0.0222  & $f_{\rm rot}$ \\
0.033  &  0.0079  &  $f_5$   \\
0.093  &  0.0052  & $f_4$  \\
\hline
\end{tabular}
\caption{
Frequencies found in the FWHM  data using a pre-whitening
procedure. }
\end{center}
\end{table}

\begin{figure}
\resizebox{\hsize}{!}{\includegraphics{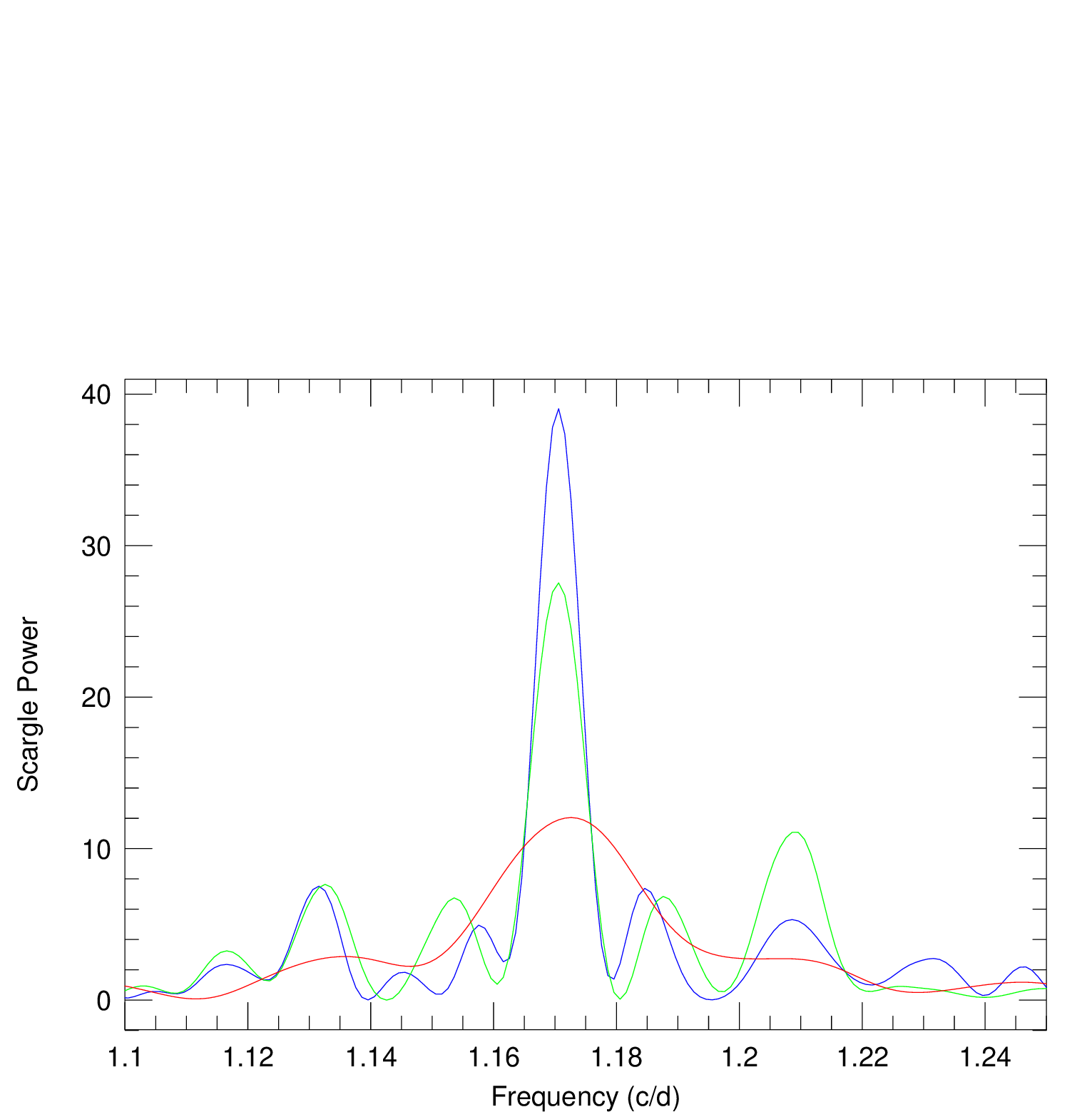}}
\caption{Scargle periodograms resulting from sequential
adding of data subsets. Red line: subset 1, green line: subset 1 plus
subset 2, blue: all data sets.
 }
\label{fapf1}
\end{figure}


\section{Statistical significance of the RV frequencies}

A common way of assessing the statistical significance of a period found
in time series data is via the Scargle periodogram (Scargle 1982). In this formulation
of the Fourier transform the power of a peak is related to a statistical
significance rather than the amplitude of the periodic signal. If
a peak in the Scargle periodogram has power, $z$, then the false alarm
probability (FAP, probability that it is due to noise) can be calculated under
two cases. The first is if one is searching for an unknown signal
over a wide frequency range, and the second is for a period known to be
present in the data. In the former, the FAP is given by 
FAP = 1 $-$ (1 $-$ e$^{-z}$)$^N$ $\approx$ $Ne^{-z}$, 
where $N$ is the  number of independent
frequencies. For the latter, the FAP is given by FAP = 
$e^{-z}$, where now there is only one independent frequency ($N=1$).

\subsection{Statistical significance of 0.85-d RV period}

Because we know that a 1.1716 c\,d$^{-1}$ signal is present in the CoRoT light
curves we should ask: {\it What is the FAP for a signal
at the known frequency of the CoRoT transit?} Table 10
 lists the Scargle power, $z$, and the false alarm
probability, FAP = $e^{-z}$ for the peak at 1.17 c\,d$^{-1}$ in each of the
data sets.  Clearly, the
0.85-d period in the RV data is significant, whereas the false alarm probability of the 
corresponding peak at the same frequency in any of the activity indicators 
is over a factor of 20 higher.

Of course, the Scargle prescription uses the 
rms scatter of the full data set to set the noise level in assessing
the FAP.  If there is real variablity in the  data this
increases the rms scatter and results in an over-estimate of the
FAP.  For weaker signals in time series that are dominated by  a stronger
one, you should subtract the contribution of the dominant signal to get
a better estimate of the FAP for the weaker signal.

\begin{table}
\begin{center}
\begin{tabular}{ccc}
\hline
Data  & Scargle  Power & FAP \\
\hline
\hline
RV     & 5.32 & 0.0049 \\
Bisector & 2.09  & 0.12 \\
Ca II  &  2.08   &  0.12  \\
\hline
\end{tabular}
\caption{Scargle power and False Alarm Probability for the Peak at 1.17 c\,d$^{-1}$, for
RV, bisector, and Ca II data. No peak  could be found near 
1.17 c\,d$^{-1}$ in the Scargle periodogram of the FWHM data.}
\end{center}
\end{table}

Figure~\ref{fapf1} shows the Scargle periodogram of the RV residuals after
subtracting all sine components {\it except} that due to the
CoRoT planet at 0.85 days.
The red line 
shows the Scargle periodogram
using Subset 1, the green is after adding Subset 2, and the blue is
the periodogram of the full data set. The fact that the Scargle power increases
with the addition of each new data set (i.e. increasing statistical
significance) gives us some reassurance that the signal is indeed present
in all data sets. The final Scargle power of the 1.171 c\,d$^{-1}$  signal  is
$z$ = 39, which results in an FAP of $\approx$ 10$^{-17}$ for 
a known signal in the data. 
The FAP that noise would
create a peak with power higher {\it anywhere}  in the full spectral
range (first equation from above)
is $\approx$ 10$^{-15}$. Thus 
this signal would be highly significant even if
we were unaware of the CoRoT transit period.

\subsection{Statistical significance of 3.7-d RV period}

The 3.7-d period ($f_2$ = 0.11 c\,d$^{-1}$) has Scargle power of 5.67. Based on this
power level we would normally not be considered the signal to be 
significant. An FAP was assessed
by using a bootstrap randomization method. The RV values of the
data (with all signals
present) were shuffled randomly keeping the observation times fixed and
a periodogram calculated for this random data. After $10^5$  such
``shuffles" the number of random periodograms having power greater
than 5.67 over the frequency range 0--0.5 d$^{-1}$) gave an estimate
of the FAP. This value was 0.3.

\begin{figure}
\resizebox{\hsize}{!}{\includegraphics{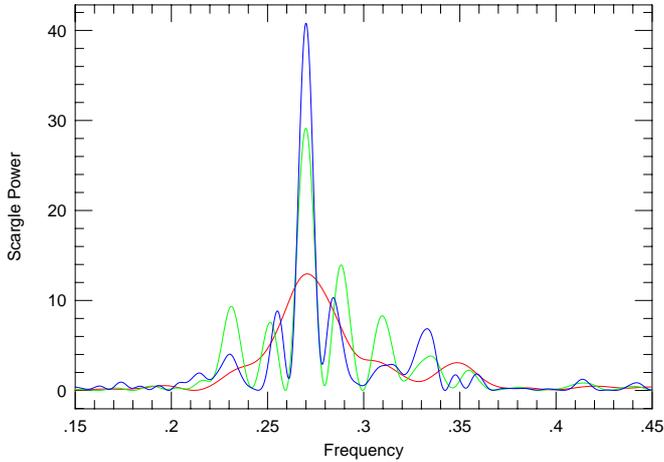}}
\caption{Scargle periodograms resulting from sequential
adding of residual RV data for the 3.7-day
period. Red line: subset 1, green line: subset 1 plus
subset 2, blue: all data sets. All sine components except the
that associated with the 9.02-day period.
 }
\label{fapf3}
\end{figure}

However, as stated earlier, this FAP may be overestimated due to the 
presence of the other signals in the RV data. To get a more realistic
assessment of the false alarm probability the 
contributions from all sine components {\it except} that due to the
3.7-day period were subtracted from the three subsets
(the 3.7-d residuals). 
Fig.~\ref{fapf3}  shows the Scargle
periodogram after sequentially adding the residual RVs from each
of the data subsets.
The Scargle power increases with the addition of each subset and the 
final value is 40  which corresponds to an FAP $\approx$ 10$^{-16}$

\subsection{Statistical significance of 9.0-d RV period}

\begin{figure}
\resizebox{\hsize}{!}{\includegraphics{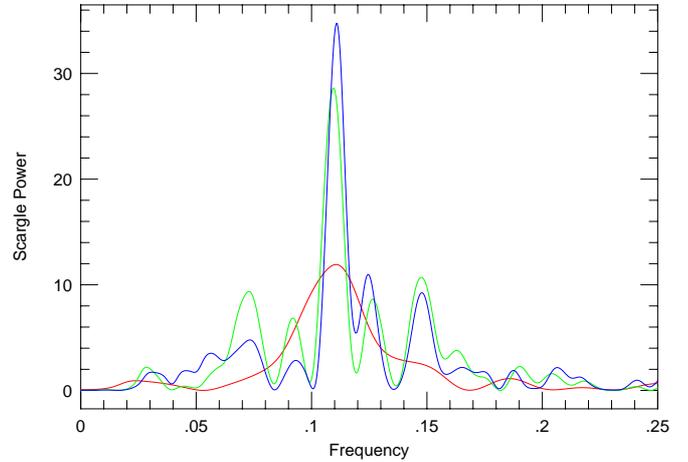}}
\caption{Scargle periodograms resulting from sequential
adding of residual RV subsets for the 9-day period. Red line: subset 1, green line: subset 1 plus
subset 2, blue: all data sets. 
 }
\label{fapf2}
\end{figure}

The peak at 9 days in the RV periodogram has Scargle power of 12.13.
A bootstrap randomization with $10^5$ shuffles yields an FAP 
$=$3.4 $\times$$ 10^{-4}$. We can be confident that this signal is
statistically significant even without subtraction of the other signals.
Fig.~\ref{fapf2}  shows the Scargle periodograms for the 9-d RV residuals, i.e. the
RV measurements with all signals except the 9-d period subtracted, and
as each subset is added. The increase in the statistical significance
with the addition of more RV measurements is an indication of a 
long-lived and coherent signal. The FAP of the final peak after including
all RV measurements is $\approx$ 10$^{-16}$

\section{Is the 0.85-d period an alias of a rotation harmonic?}

The Scargle periodograms in Fig.~\ref{scargles} are 
dominated by the rotational frequency, $f_{\rm rot}$. However, there is also power at the first
two harmonics of the rotational frequency, namely 2$f_{\rm rot}$ and  $3f_{\rm rot}$. Indeed, Q09 applied harmonic filtering using 
the rotational frequency and its first two harmonics to remove the signature of activity. Because of this there
is some concern that the 1.1715 transit frequency is close to the one-day alias of the 3rd harmonic
of the rotational frequency (i.e.  $4f_{\rm rot}$ $+1$ = 1.1715 c\,d$^{-1}$).  In the best case, the activity signal can thus contribute
some power at the orbital frequency of CoRoT-7b. 
In the worse case all the RV signal at this frequency could arise from
activity.

Q09  argued that because repeated measurements were taken on 
the same nights, that this alias effect was minimized and that the 
0.85-d RV period was the one actually present in the data. To investigate whether alias
effects of a rotational harmonic can account for the observed 0.85-d period in the RV data
 we performed an analysis on a subset of the RV consisting  only of those nights for which 
at least 3 RV measurements were made during the night. In the following analysis it is assumed that the
activity signal contributes a constant value to the measured RV on a given night. This is a reasonable
assumption. The maximum time separation between the first and last measurement on a given night is
less than 4 hours. This corresponds to a change in rotation phase of only 0.007. The contribution of stellar
activity to the measured RV should thus be a constant over that time. This also assumes 
that spot evolution  over 4 hours is negligible. Meanwhile, the
change in orbital phase of CoRoT-7b is 0.2 and should cause the dominant variations  observed in
the RV during the course of the night.

It is not strictly true that the non-CoRoT-7b RV variations are constant during the night. The orbital motion due
to the CoRoT-7c should produce  an appreciable variation over 4 hours. However, this contribution
 is also relatively small. Over 4 hour time span the orbital phase of CoRoT-7c only changes by 0.05 which corresponds to a maximum RV
variation of only 1.5 m\,s$^{-1}$ well below the measurement error.

There were 7 nights where at least 3 RV measurements of CoRoT-7 were made 
resulting in  a
total of 21 measurements. The data from these nights were treated as 7 independent data sets with each
one have a different a  zero-point velocity which could vary  from night to night. A least squares
sine-fit the 7 data sets was made keeping the 
period fixed to the transit period of 0.8535 days, but allowing phase and the zero-point value for  each night to vary.
The advantage of such an approach is
that it is a ``minimal impact'', low pass filter that makes no assumptions about the underlying time variability
due to stellar  activity, only that it is constant on a given night.

\begin{figure}
\resizebox{\hsize}{!}{\includegraphics{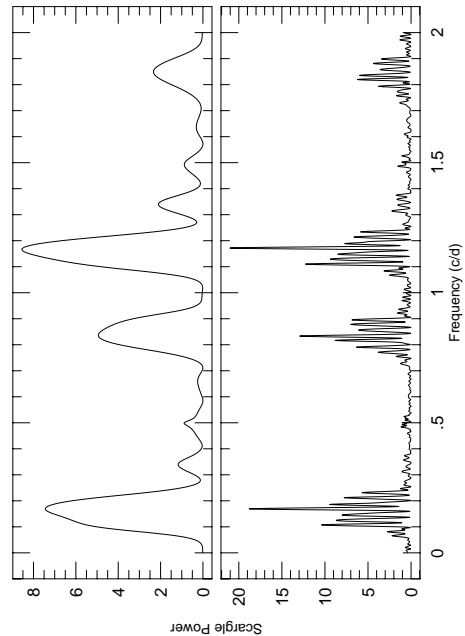}}
\caption{(Top) Scargle periodogram of the data set consisting of 7 nights
where at least 3 RV measurements were made. The measurements from each night
were  considered an independent data set and a fit was made using the
period of CoRoT-7b and allowing zero point offset from each night to vary.
The periodogram is of the data with the offset to each night subtracted.
(Bottom) Same as in the top panel but with the addition of 16 nights for which
two RV measurements were made per night. 
 }
\label{ftshort}
\end{figure}

\begin{figure}
\resizebox{\hsize}{!}{\includegraphics{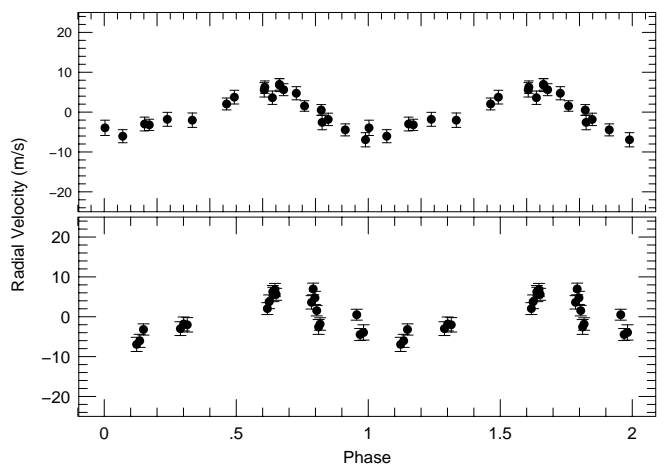}}
\caption{(Top) RV measurments of the 7-night data set (zero-point offsets 
applied) phased to the period of CoRoT-7b. (Bottom) same measurements
phased to the alias period of 5.82 days.
}
\label{phasecompare}
\end{figure}

The top panel of Figure~\ref{ftshort} shows the  
Scargle periodogram of these 21 RV measurements after
subtracting the individual zero-point offsets determined by the least squares fitting. The
strongest peak is at the transit frequency of 1.1715 c\,d$^{-1}$ and 
with a smaller peak at the
alias frequency of 0.1715 c\,d$^{-1}$. Figure~\ref{phasecompare} 
shows the RV measurements
phased to the period of CoRoT-7b  and the alias period (5.7 days). The ``cleaner'' phase diagram
of the 0.85 d period supports that this is the true period in the data, and not the alias of 
$4f_{\rm rot}$.

The false alarm probablity of the peak at 1.1715 c\,d$^{-1}$ was
assessed using a bootstrap randomization procedure. The RV values  were shuffled
2 $\times$ 10$^5$ times keeping the observed times fixed and periodograms calculated
for this random data. Over the frequency range 0 $<$ $\nu$ $<$ 2 c\,d$^{-1}$
there was only four instances
where the periodogram of the random exceeded the power of the real data. 
The FAP for this signal is thus $\approx$ 2 $\times$ 10$^{-5}$. We should note that this
bootstrap was evaluated over the full frequency range 0--2 c\,d$^{-1}$. 
Since we are interested in the known frequency of CoRoT-7c it is more
appropriate to evaluate the bootstrap
over a much narrower range centered on $\nu$ =
1.1715 c\,d$^{-1}$. The FAP of this signal is almost certainly much less.

There were an additional 16 nights where 2 RV measurements were taken 
of CoRoT-7.  The time separation of these points are large enough to provide
some good sampling of the 0.85-d sine wave presumed to be in the data.
These nights were added to the data subset of 3 points per night 
to give a total of 53 data points spread over 23 nights. 
A new fit was performed, again allowing the 
nightly zero-point  values to vary, but keeping the period fixed to the transit period.
 The lower panel of Fig~\ref{ftshort}
shows the periodogram of these data with the zero point offsets applied. Note
that the Scargle power has significantly increased indicating a more signficant
detection (FAP $\approx$ 10$^{-16}$). Also, the  frequency at 
1.1716 c\,d$^{-1}$ is still higher than the alias frequency. A phase diagram
of these data will be shown below when we present orbital solutions.

\begin{figure}
\resizebox{\hsize}{!}{\includegraphics{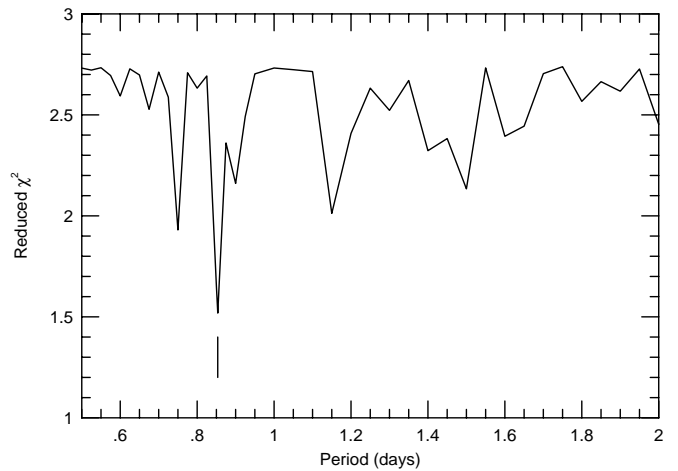}}
\caption{The reduced $\chi^2$ as a function of period for the data set
consisiting of nights with multiple RV measurements. The zero-point offset
from each night was allowed to vary. The vertical line indicates the period
of CoRoT-7b.
}
\label{redchi}
\end{figure}

We also investigated whether the period of CoRoT-7b could be extracted from the
data without a priori knowledge of the transit period. 
To do this a range of periods
were fit to the 53 RV measurements of the data set with 
multiple observations per night. At each trial fit the period was fixed, but the phase and 
zero-point offset were allowed to vary. Figure~\ref{redchi} shows the reduced
$\chi^2$ fit to the data as a function of the fitting period. Good fits 
occur  at 
periods where $\chi^2$ is minimized. This procedure  could be considered as a variant
of the phase dispersion minimization technique of Stellingwerf (1978), but 
which allows the mean value of
individual data sets to float (i.e. a ``floating mean'' phase dispersion minimization). The 
deepest $\chi^2$ minimum is near the period of CoRoT-7b (vertical line).
Other minima are clearly alias periods and phasing of the data to these periods 
does not produce as clean a phase diagram as for the CoRoT-7b period.

\subsection{Fourier analysis of nightly offsets}

\begin{figure}
\resizebox{\hsize}{!}{\includegraphics{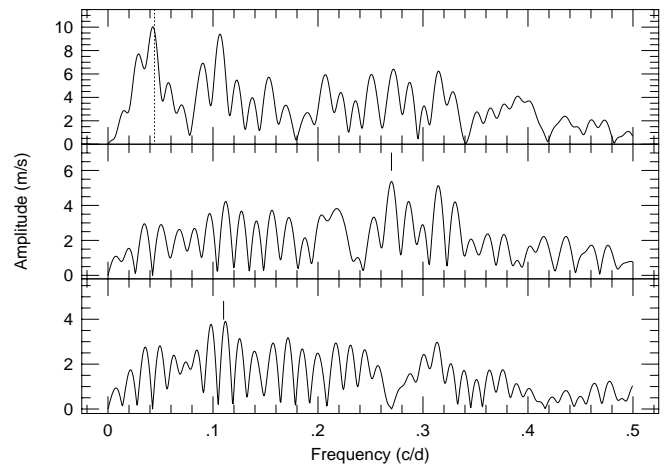}}
\caption{A pre-whitening on the zero-point offsets of the multi-measurement
nights found by the least squares fitting. (Top) Amplitude spectrum of the
zero-point values. The dashed line indicates the rotation period of CoRoT-7b.
(Middle) The amplitude spectrum after removing the rotational frequency.
The vertical line corresponds the frequency of the 3.7-d period of CoRoT-7c.
(Bottom) The amplitude spectrum after also removing 3.7-d period signal.
The vertical line corresponds to the frequency $f_3$.
 }
\label{offset}
\end{figure}

As a sanity check we performed
a Fourier analysis on the nightly offsets produced by our least  
squares fitting. If these derived nightly offsets have some relationship
to the activity we should see evidence of the rotational period. 
 The top panel of Fig.~\ref{offset} shows the Fourier
transform of the nighty offsets (23 values). The highest peak corresponds to the rotational 
frequency of CoRoT-7 (indicated by the vertical dashed line). This indicates that the
nightly zero-point offset indeed follows the RV rotational modulation due to activity.

Of interest is when we pre-whiten the nightly offsets to search for additional periods.
The second panel shows the offset data after removing the contribution of the rotational
frequency. The highest peak corresponds to a frequency of 0.27 c\,d$^{-1}$ (indicated by the vertical 
line). The bottom panel shows the amplitude spectrum of the residuals after removing
the contribution of the 0.27 c\,d$^{-1}$ frequency. The highest peak 
corresponds to 
0.112 c\,d$^{-1}$, close to the
0.11 c\,d$^{-1}$ frequency found in the RV (indicated by the vertical line). It is reassuring that the 
nightly offsets show evidence for the rotation period of the star, but also the 3.7-d and 9-d
period found in the RV analysis.

\section{On the nature of the 9-d period}

In Q09  the two approaches to the RV data analysis (harmonic filtering and
 Fourier
pre-whitening) agree on the presence of a 0.85-d and 3.7-d period in the data.
The main difference to the two approaches was that the pre-whitening procedure
yielded a significant period at 9-d, whereas this signal was absent in the harmonic filtering of activity 
signal. The obvious interpretation is that the 9-d period arises from the activity signal.
However, the harmonic filtering worked on sub-sets of the data that were roughly
the length of the rotation period, whereas the pre-whitening
procedure was performed on the full dataset.
 The 9-d period is very close to $P_{\rm rot}/2$, i.e.
the first harmonic of the rotation period. There is a danger that when applying
harmonic filtering one may remove real signals not associated with a rotational harmonic.

\begin{figure}
\resizebox{\hsize}{!}{\includegraphics{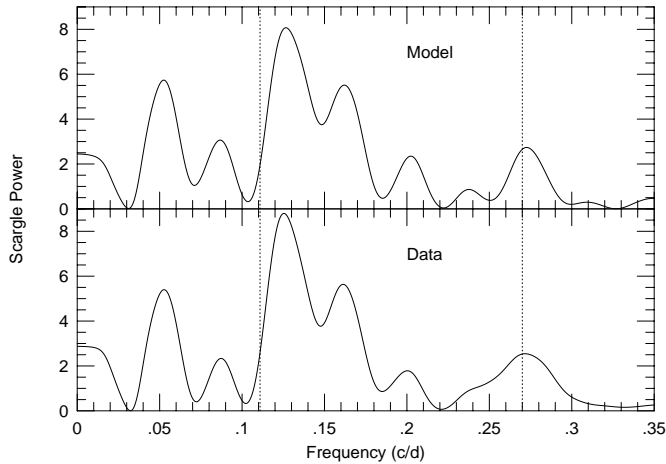}}
\caption{
(Top) Scargle periodogram for the fit to RV first data subset  (JD=4775-4807)
sampled in the same manner as the data and with noise at a level of
$\sigma$ = 2 m/s. (bottom) The Scargle periodogram of the real data.
The vertical lines mark the frequencies of the 9-d and 3.7-d signals that
are in the fake data.
 }
\label{sim1}
\end{figure}

To test this hypothesis the frequency solution of the Subset 1 (Table 3) 
was used
to generate a fake data set. This fake data included the signals $f_1$, $f_2$,
and $f_3$.
The fits were sampled in exactly the same way as the real data and 
noise at a level of $\sigma$ = 2 m\,s$^{-1}$ was also added. The Scargle periodogram of the
fake data (top panel) is  compared to the periodograms of the real data
(lower panel) in Figure~\ref{sim1}.  The frequency of the input
3.7-d and 9-d period signals are shown as vertical lines.

There are two things to note about this figure. 
The periodogram of the fake data looks
exactly like the real  data, as it should. After all, the rms scatter
about the fit is under 2 m\,s$^{-1}$. If you fit the data you fit the periodogram.
 Note, however, because of the short data window the true frequency in the
 data, $f_3$, appears at a slightly different frequency of 0.12 c\,d$^{-1}$.

\begin{table}
\begin{center}
\begin{tabular}{cc}
\hline
Frequency   &  Amplitude   \\
(c\,d$^{-1}$)        &       (m\,s$^{-1}$) \\
\hline
\hline
0.044  &  11.9   \\
0.088  &    5.8       \\
0.132  &  4.5     \\
0.022  &  4.6            \\
\hline
0.270   &  6.02               \\
1.1716  &  4.4               \\
\hline
$\sigma$ = 1.99 m\,s$^{-1}$ \\
\end{tabular}
\caption{
Frequencies found in the simulated data set of Subset 1
(Fig.~\ref{sim1}) using a pre-whitening procedure and harmonic
analysis. The data was first fit using the rotational harmonics
above the horizontal line at fixed frequency, but varying amplitudes.
The frequencies below the horizontal line are additional frequencies
found by the pre-whitening procedure.
}
\end{center}
\end{table}

\begin{table}
\begin{center}
\begin{tabular}{ccc}
\hline
Frequency   &  Amplitude &   Comment  \\
(c\,d$^{-1}$)        &       (m\,s$^{-1}$)&  \\
\hline
\hline
0.0450  &  10.46  & $f_{\rm rot}$ \\
0.0841  &  4.27  &           \\
0.1384   &  2.43 &            \\
0.1093   &  6.01 &                \\
0.2701    & 5.51 & $f_2$          \\
0.0945    &  6.86 &   \\
1.1706    &  4.50 & $f_{3}$ \\
0.0357   &  4.64& $f_{1}$    \\
0.0728   & 2.33  &                    \\
\hline
\end{tabular}
\caption{Pre-whitened frequencies found in the RV data after first fitting
the rotational frequency and its first 2 harmonics. The final rms
about the fit is 2.35 m\,s$^{-1}$}
\end{center}
\end{table}

This fake Subset 1 data set was analyzed using harmonic analysis
and pre-whitening. The data was first fit using the rotational frequency
and its first two harmonics ($f_{\rm rot}$, 2$f_{\rm rot}$, and 3$f_{\rm rot}$) as well
as 0.5$f_{\rm rot}$ (there was evidence for the presence of this frequency,
see Table 1.). The frequencies were kept fixed, but the amplitude and phase
were made to vary. The pre-whitening procedure was then carried out to 
find additional frequencies. The results are listed in Table 11. The
pre-whitening procedure found only 2 additional frequencies, $f_1$ and
$f_2$, but not $f_3$ even though it was present in this fake data.
The same result was found when applying this procedure to the other
data sets.

Harmonic filtering was also performed on the full RV (real) data set
by first fitting $f_{\rm rot}$, 2$f_{\rm rot}$, and 3$f_{\rm rot}$ to the data and 
then continuing the pre-whitening procedure to find additional frequencies.
After all frequencies were found the solution optimized by fitting all
components simultaneously.
The results are shown in
Table 12. The frequencies $f_1$, 
$f_2$, and $f_3$ were recovered in spite of the harmonic pre-filtering. 

This investigation  has demonstrated that one should be careful in filtering the time
series data using  harmonics of the
rotational frequency of CoRoT-7 on a limited time span data set. In this case,
using rotational harmonics essentially filtered out the
real period that was in the simulated data.
 When applying rotational harmonic
filtering to the full real data set the 9-d period was still recovered. 
 The reason for this is that over a 23-d time span there is
little  difference between $\nu$ = 0.11 c\,d$^{-1}$ and the first harmonic
of the rotational frequency, 
$\nu$ = 0.09 c\,d$^{-1}$. Harmonic filtering will essentially  remove this signal. However,
over the 100 days that the RV data were acquired the subtle frequency difference
between $f_3$ and 2$f_{\rm rot}$ can be resolved and harmonic filtering cannot
fully remove the signal due to $f_3$.  When performing an analysis on data subsets
it is instructive to 
look also at the full data covering the longest time span.

\subsection{Bisector $-$ RV correlations}

The HARPS data also contain three indicators of activity:
the FWHM of the CCF, the bisector of the CCF, and the Ca II
emission measure. 
Of these 3 activity indicators only the line
bisectors have a direct relationship to the RV variations due to
activity. Ca II emission originates in plage and these regions do 
not necessarily have the same surface distribution as spots. For
cool spots the FWHM mimics the photometric variations. The FWHM is
a minimum (like the photometry) when the spot distortions are in the
wings of the spectral line (i.e. limb of star) and a maximum when the
spots are at disk center. Thus there is a phase shift of $\approx$
$P_{\rm rot}/4$ between the maximum of the RV and FWHM. During this
time spot evolution may  be significant.  On the other hand, surface
spots produce a distortion in the spectral line which results in a shift
in the line centroid. Thus any RV variations due to activity must be 
correlated {\em directly} with variations in the spectral line bisector without the
need to apply any phase shift.  For this reason we
will focus on the RV - bisector variations in CoRoT-7 to assess the nature
of the 9-d period.

\begin{figure}
\resizebox{\hsize}{!}{\includegraphics{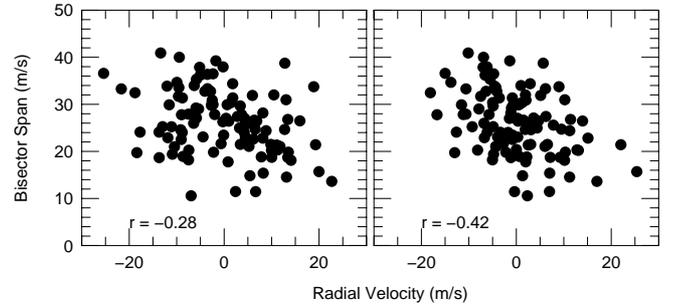}}
\caption[]
{(Left) Bisctor span of the CCF versus the RV measurements. 
The correlation coefficient is $-$0.28. (Right) The Bisector-RV
correlation after $f_1$, $f_2$, $f_3$ (planet signals) have been
removed from the data. The correlation coefficient is $-$0.42.
}
\label{biscorr}
\end{figure}

RV variations caused by cool spots should show an anti-correlation 
(negative slope) with the bisector variations (see Queloz et al. 2001). 
The RV variations of CoRoT-7 do show a slight anti-correlation with the
bisectors (left panel of  Fig.~\ref{biscorr}). 
If an RV signal is not due to activity, then removing this from the
observed RV measurements should result in  a stronger correlation
between the bisectors and RV variations.  Indeed, when one removes that signal
of the 0.85-d, 3.7-d, and 9-d period from the RV data the bisector span - 
RV variations become more correlated ($r$ = $-$0.42, right panel). This suggests
that the 0.85-d, 3.7-d, and 9-d periods (i.e. $f_1$, $f_2$, and $f_3$) found in the RV data are not
associated with activity. 

\begin{figure}
\resizebox{\hsize}{!}{\includegraphics{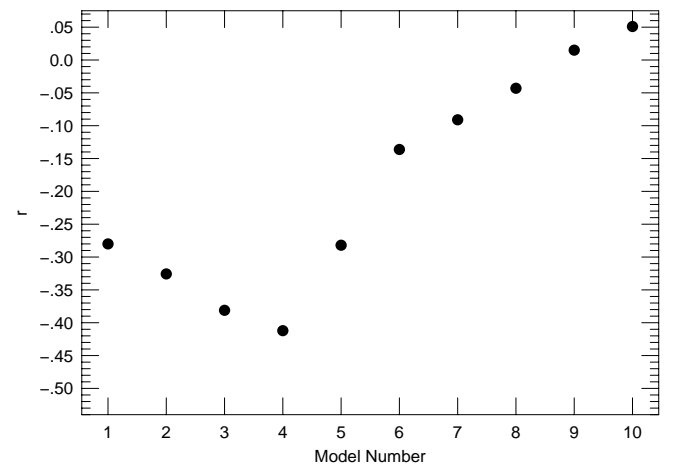}}
\caption[]
{The correlation coefficient of the CCF bisector span with RV
as a function of different versions of the RV data with the
frequencies in Table 1 removed. Model 1: raw
data, Model 2: $f_1$ removed, Model 3: $f_1$ and $f_2$ removed,
Model 4: $f_1$, $f_2$, and $f_3$ removed. Model 5: Model 4 and
$f_{\rm rot}$ removed. Others: subsequent removal of all other frequencies
in Table 1.
}
\label{corrtrend}
\end{figure}

Fig.~\ref{corrtrend} shows the correlation coefficient between the
bisector and RV measurements as a function of data sets with the contributions
of various frequencies removed (denoted ``Model Number'')
Model 1 is to the
original RV data set. Model 2 is this data set with the 0.85-d period removed.
Model 3 is the previous model,  but with the 3.7 d period also removed.
Model 4 is the data set with the 0.85-d, 3.7-d, and 9-d periods removed. Note
that the correlation coefficient becomes more negative with the removal of each
data set suggesting that these 3 periods are not associated with stellar activity.

Model 5 represents Model 4 $-$ $f_{\rm rot}$. Each subsequent model is the previous
model with
successive frequencies listed in Table 1 removed (and skipping of course
$f_1$, $f_2$, and $f_3$ already subtracted in Model 4). The fact that the
RV-Bisector variations become more uncorrelated with 
the removal of additional frequencies
suggests that all frequencies in Table 1 except $f_1$, $f_2$, and
$f_3$ are most likely due
to activity.

Because the RV variations are directly related to the bisector
variability we can use the temporal variations of the RV to produce
a predicted bisector variability. Saar \& Donahue (1997) gave relationships
relating both the bisector span and RV amplitude as a function of spot filling
factor and rotational velocity of the star. The RV-to-Bisector amplitudes
ratio from their expressions is predicted to be about a factor of 10. However,
the exact ratio depends on several factors, primarily how one measures
the bisector span and which spectral lines are used. Their result cannot
be directly compared to the bisector measurements of the CCF from the HARPS
data.
A better way is to use the  RV-to-Bisector amplitude 
ratio estimated by comparing the amplitudes of 
frequencies found in both the  RV  and bisector amplitude spectra and using data
where these quantities (bisector and RV)  were measured in a consistent way.
The RV frequencies $\nu$ = 0.043 c\,d$^{-1}$ and 
0.094 c\,d$^{-1}$ in Table 1 have amplitudes of 7.5 m\,s$^{-1}$ and
5.69  m\,s$^{-1}$. The corresponding frequencies in the bisector amplitude
spectrum 
have amplitudes of 4.21  m\,s$^{-1}$ and 2.81  m\,s$^{-1}$, respectively.
This implies that the RV-to-Bisector  amplitude is $\approx$ 2.

We created a  model of the bisector
variations based on the amplitude spectrum of the RV measurements.
Two fake data sets of bisector variations were generated. The first used  all 
the frequencies in Table 1, excluding  $f_1$, $f_2$, and $f_3$.
All  amplitudes in Table 1 were
reduced by a factor 0.5 corresponding to the bisector-to-RV amplitude ratio.
 The sampling of this time series was the
same as the actual data. The bisector variations have an rms scatter
of 4.5 m\,s$^{-1}$ after removal of the dominant frequencies. This was taken
as the mean bisector error and random noise at this level was added to the
fake bisector data.  The second data set was similar to the first one, but
the frequency $f_3$ was present with the appropriately scaled amplitude
(i.e. only $f_1$ and $f_2$ were removed).

\begin{figure}
\resizebox{\hsize}{!}{\includegraphics{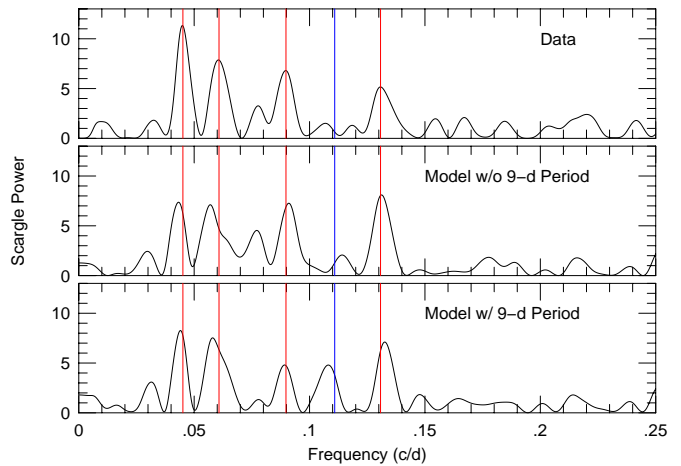}}
\caption[]
{(Top) Scargle periodogram of the CCF bisector variations.
(Middle) Scargle periodogram of a fit generated using the
RV frequencies of Table 1, but without $f_1$, $f_2$, and $f_3$
present. Amplitudes were adjusted according to the measured
RV-to-Bisector amplitudes. Random noise at the level observed
for the bisector measurements were also added. (Bottom) Same
as for the middle panel but with $f_3$ (9-d period) present.
}
\label{rvbismodel}
\end{figure}

The top panel of Fig.~\ref{rvbismodel} shows the Scargle periodogram of
the actual bisector measurements. The middle panel shows the periodogram
of the fake data, but without the presence of the 9-d period. The bottom panel 
shows the periodogram of the fake bisector data with the 9-d period present.
Vertical red lines show the rotation frequency and its first 3 harmonics. The vertical
blue line shows the frequency corresponding to the 9-d period.

There are two interesting features about this figure. 
First, the periodogram of the
fake bisector data without the 9-d period looks qualitatively like the real data.
Many of the same peaks are seen in both periodograms.
The ratio of the amplitudes are not quite correct probably due to 
the bisector data having more complicated noise characteristics than the 
Gaussian noise in our simple model. Furthermore, for 
high frequency components that the RV-to-Bisector amplitude ratio may be
different. The second important feature to note is that in the  periodogram of the
fake data with the 9-d period present
there is a signficant peak at $\nu$ = 0.11 c\,d$^{-1}$ 
that is not seen in the data periodogram. If the 9-d period was 
due to activity we should have seen a 
corresponding peak in the periodogram of the bisectors and most likely
in the Ca II and FWHM.
This also argues in favor
of the 9-d period not being related to activity.

\section{A search for transits from CoRoT-7d}

The CoRoT-7 light curve was analyzed to see if a 9-d transit signal 
could be found in the data. A similar investigation  was already performed 
to search for transits from CoRoT-7c, but none was found (Q09).
If the orbits of all planets are co-planar we do not expect to see transits from CoRoT-7d. 
However, if substantial differences of the orbital inclination between the planets exist,
CoRoT-7d may well transit in spite of having a larger orbital radius than CoRoT-7c.

The CoRoT-7 light curve first had to be filtered for the large variations due to 
the stellar activity.  This was done by: 1) normalizing the light curve by the maximum value,
2) performing outlier rejection, 3) reducing residual orbital effects by using a running median
of length one orbit, and 4) tracking slow variations using a sliding polynomial of
12 hours and subtracting these. Finally, we fit the transit curve of CoRoT-7b and
subtracted that from the light curve.

This processed light curve was then phase-folded to the 9.02 d period. No transit
signal above the noise was detected.  We derive an upper limit of 1.5 $\times$ $10^{-4}$
for the depth of any transit at a period of 9.02 d.

\section{Orbital solutions}

\subsection{The 0.85-d period}

Models of the internal structure of CoRoT-7b rely on the mass determination
which in turn hinges on the amplitude of the RV variations.
In Q09 the amplitude of CoRoT-7b was estimated to be 3.5 m\,s$^{-1}$
using two different approaches and this
corresponds to a planet mass of 4.8 $M_{\oplus}$.   However, this value
was obtained after applying a correction term for the  effects of the filtering
process to both techniques using simulated data.
The uncorrected pre-whitening
procedure amplitude was slightly higher at  4.2 m\,s$^{-1}$ and
the harmonic filtering amplitude lower at 1.9 m\,s$^{-1}$. 
Clearly, 
the amplitude of CoRoT-7b depends on how one removes the activity signal.

As an alternative approach to determining the RV amplitude of CoRoT-7b we
took the results from Section \S6. In this analysis
we used only the RV data for which
multiple measurements were made each night. A least squares sine 
fit to this data was made keeping the period fixed and allowing the nightly
offset to vary. The final offsets were then applied to the individual
nights and the data combined.
This may be the best way to account
for the RV variations of activity without any assumptions about its
temporal behavior.
An orbital solution was performed on all the residual RVs keeping the
ephemeris, $T_0$,  fixed to the CoRoT transit time 
of 2454446.7311. 

\begin{figure}
\resizebox{\hsize}{!}{\includegraphics{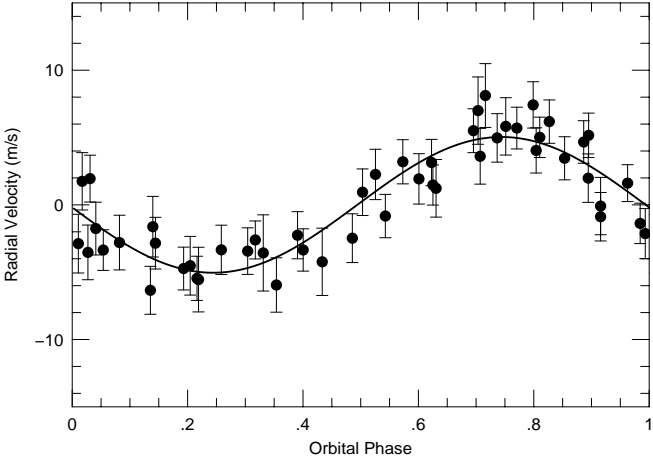}}
\caption{(Top) Orbital solution for the 0.85-d period using the data
with repeated nightly measurements and the appropriate zero-point offset
applied.
 }
\label{orbit7b}
\end{figure}

Table 13  lists the orbital elements. 
At first the nightly data were fit keeping the CoRoT transit period
of 0.8535 days. If we allow this parameter to vary we get a best fit to the
data with a period of 0.85359 days which is listed in the
table.  Allowing the $T_o$ to vary, but keeping the period fixed results
in  $T_o$ = 2454446.7330 a value  very close to the CoRoT ephemeris.
Allowing the eccentricity to vary results in a best fit value of 0.08, but
with large error, $\sigma$ =  0.13.
We cannot exclude a slight eccentricity in the orbit, but given the 
large variations due to activity and the additional companions, this may
be difficult to extract reliably from the RV data.
Figure~\ref{orbit7b} shows the zero-point corrected data phased to the
CoRoT transit ephemeris and a period of 0.85359. The solid line
represents the orbital solution.

The derived K-amplitude is 5.04 $\pm$ 1.09 m\,s$^{-1}$ which
results in a companion mass of 6.9 $\pm$ 1.4 $M_{\oplus}$. This is
slightly larger than the K-amplitude of 4.16 $\pm$ 0.27  m\,s$^{-1}$
($m$ = 5.75 $\pm$ 0.37  $M_{\oplus}$) by pre-whitening the full data set
(Q09). The $K$-amplitude from pre-whitening of the 
full data set and the analysis of the subset RV data with
multiple measurements each night  both suggest a slightly higher planet mass
than the 4.8 $\pm$ 0.8 $M_{\oplus}$ of Q09, although all determinations
are consistent to within the errors.

\begin{table}
\begin{center}
\begin{tabular}{lr}
\hline
Parameter  & Value  \\
\hline
\hline
Period [days]  &  0.85359  $\pm$ 0.00059\\
T$_{0}$ [JD] (fixed) & 2454446.7311 \\
$K$ [m\,s$^{-1}$] & 5.04 $\pm$  1.09 \\
$e$ (fixed)              & 0.00  \\
$m$  [$M_{\oplus}$]       & 6.9 $\pm$ 1.4\\
$a$ [AU] & 0.017 \\
\hline
\end{tabular}
\caption{Orbital parameters for the 0.85-d period.
}
\end{center}
\end{table}

\begin{table}
\begin{center}
\begin{tabular}{lr}
\hline
Parameter  & Value  \\
\hline
\hline
Period [days]  &  3.691 $\pm$ 0.0036\\
T$_{\rm periastron}$ [JD] & 2454450.38 $\pm$   0.03 \\
$K$ [m\,s$^{-1}$] & 5.10 $\pm$  0.33\\
$e$               & 0.08   $\pm$ 0.05  \\
$\omega$ [deg]    & 166 $\pm$ 36 \\
$m$  [$M_{\oplus}$]       & 12.4 $\pm$ 0.42\\
$a$ [AU] & 0.045 \\
reduced $\chi^2$  & 1.07  \\
\hline
\end{tabular}
\caption{Orbital parameters for the 3.7-d period. 
}
\end{center}
\end{table}

\begin{table}
\begin{center}
\begin{tabular}{lr}
\hline
Parameter  & Value  \\
\hline
\hline
Period [days]  &  9.021 $\pm$ 0.019\\
T$_{\rm periastron}$ [JD] & 2454446.16 $\pm$   0.09 \\
$K$ [m\,s$^{-1}$] & 5.90 $\pm$  0.83\\
$e$ (fixed)              & 0.0  $\pm$ 0.05 \\
$m$  [$M_{\oplus}$]       & 16.7 $\pm$ 0.42\\
$a$ [AU] & 0.08 \\
reduced $\chi^2$  & 1.24  \\
\hline
\end{tabular}
\caption{Orbital parameters for the 9.02-d period. 
}
\end{center}
\end{table}

\subsection{The 3.7-d period}

RV residuals were produced by subtracting
all sine components except for $f_2$ from the full RV data
set and an orbital solution calculated. In removing the
contribution of $f_1$ the amplitude in Table 13 was used rather than the slightly higher 
amplitude found in Table 1.  The
derived amplitude is slightly lower than the one presented in Q09 (5.5 m\,s$^{-1}$) that
removed $f_1$ using the amplitude found in the pre-whitening procedure. In order to
estimate the range of velocity amplitudes for $f_2$ a least squares fit to the original RV 
data was first made using this frequency and the subsequent frequencies found in the
pre-whitening process sequentially subtracted.  The range of amplitudes for $f_2$ during this
process ranged from 4.75 to 5.4 m\,s$^{-1}$. An error of  $\sigma$ =  0.33   m\,s$^{-1}$ was adopted  which is
slightly more than the formal error of  $\sigma$ =   0.39   m\,s$^{-1}$ from the orbit fitting. Note that this formal error is much lower than for
the amplitude CoRoT-7b which was calculated using only a subset of the data.

\begin{figure}
\resizebox{\hsize}{!}{\includegraphics{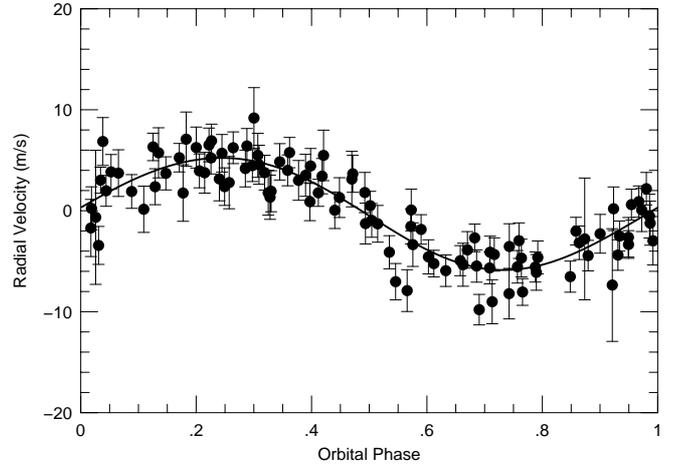}}
\caption{ The RV residuals of CoRoT-7 after removing all components in
Table 1 except for $f_2$ and 
phased to the 3.7 d period (points). The orbital fit is shown as a line
}
\label{orbitf2}
\end{figure}

Fig.~\ref{orbitf2} shows the orbital solution the 3.7-day period. The orbital elements
are listed in Table 14. A slight eccentricity is found, but this may well
be an artifact due to the filtering process.

\subsection{The 9-d period}

The RV residuals were calculated subtracting
all sine components in Table 1 except for $f_3$ and again using the
amplitude for $f_1$ from Table 13. The range of possible amplitudes for 
$f_3$ during the pre-whitening procedure was 5.5 -- 7.15 m\,s$^{-1}$.
We thus adopted an error in the ampitude of $\sigma$ = 0.83  m\,s$^{-1}$,
larger than the formal error of  $\sigma$ =  0.26  m\,s$^{-1}$, but probably a
more realistic estimate. A best fit orbit allowing the eccentricity
to vary resulted in a slightly negative value. We therefore took a solution with 
the eccentricity fixed to zero with an error of   $\sigma$ = 0.05.

\begin{figure}
\vspace{3.0in}
\resizebox{\hsize}{!}{\includegraphics{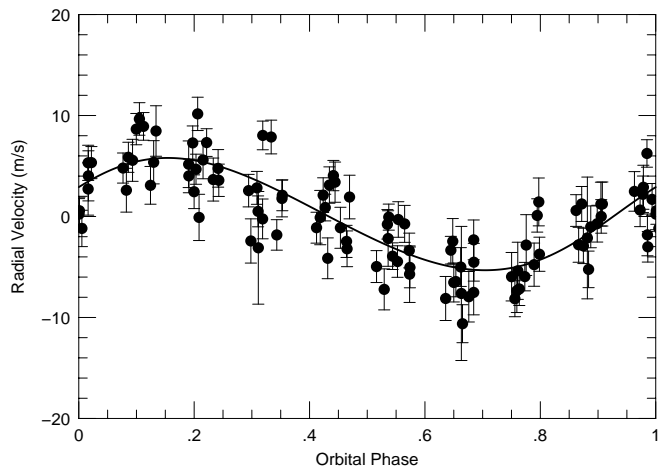}}
\caption{The RV residuals of CoRoT-7 after removing all components in
Table 1 except for $f_3$ and 
phased to the 9.02 d period (points). The orbital fit is shown as a line
}
\label{orbitf3}
\end{figure}

Fig.~\ref{orbitf3} shows the orbital solution to the RV residuals using the
9-day period. 
The orbital elements are listed in Table 15.  Note the gap-like structures
in the phase curve. This is undoubtedly due to the period being nearly
an integer value of one day, i.e. our typical sampling rate.

\section{On the dynamical stablity of the 3-planet system.}

We also investigated whether the 3-planet system would be 
dynamically stable. Although a stable system is no proof that all
3 planets exist, an unstable system would indicate 
that the additional planetary signals found in the HARPS data
may arise from activity. 

\subsection{An ultra-compact planet system?}

The stability of planetary systems involves multiple interactions of planets by
mutual perturbations
often involving interactions of resonances between more than two planets and various
components of 
their motions. It is a stability for some time. As Lecar et al. 2001
summarize it: 
{\em The solar system is not stable, it is just old!}. Given the snapshot provided
by discovery orbital elements 
we first briefly overview the gross/overall properties of the CoRoT-7 system by a
simple stability 
indicator based on Hill-exclusion volumes. It requires non-overlapping cylindrical
volumes 
of half-width $k R_{\rm Hill}$, around each planet orbit. The Hill-radius to lowest
order is 
$R_{{\rm Hill},i}=k a_i (m_i/3{\rm M_\odot})^{1/3}$, with
$m_i,a_i$ denoting the mass and semi-major-axis of the $i$\,-th planet, 
respectively, and $k$ is a factor of 4--15 depending on the number of planets, the
dynamical structure of 
the system (masses, orbital elements), and the time-scale for stability under
consideration, c.f.\ 
e.g.~Chambers et al. (1996) including a discussion on system stability with
long-term orbital
calculations. 
Funk et al. (2010) show that for close-in systems, there are surprisingly large
volumes of phase 
space for stable and planet-rich systems at periods of less than 10 days around
solar mass stars. These {\em ultra compact systems} can easily harbor 8 {\em Super Neptunes} for 
$500\,{\rm Ga}$ and up to $30\,{\rm M_{\oplus}}$.  Funk et al. (2010)
determine a
factor of $k \ga 6.4$ that is necessary for stability in the present context --- 
3 planets, masses below Neptune's and few Ga timescales --- and give
examples for it being sufficient including close-in systems with an 
additional Jupiter-mass planet at $0.26 {\rm AU}$ (i.e.\ $\sim 50$ days).

\begin{figure*}[htbp]
  \resizebox{\hsize}{!}{\includegraphics{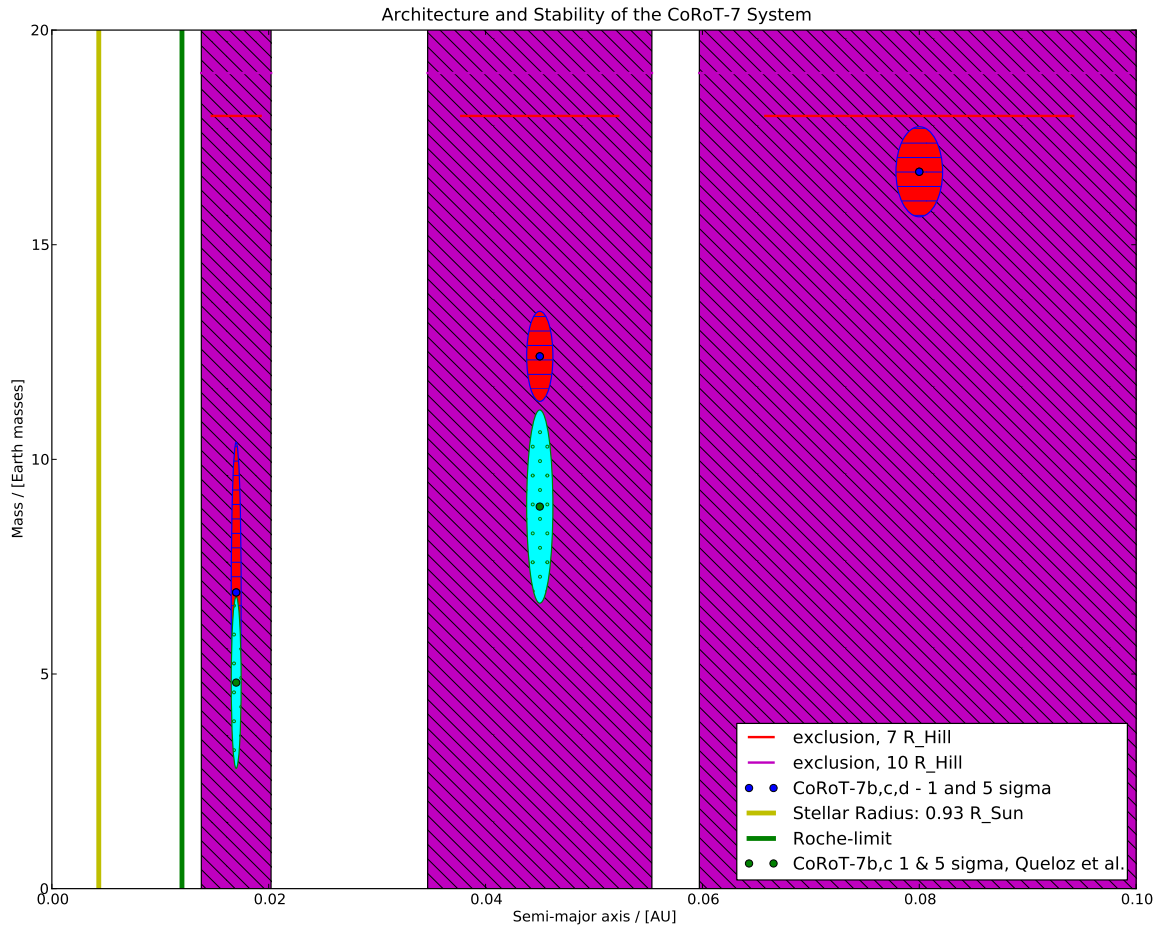}}
  \caption{Synopsis and simple stability for the CoRoT-7 system. Planetary mass in
Earth-units
  is plotted against the distance from the star in AU. Mass and semi-major axis
determinations 
  of this work (blue dots) and the discovery paper (green dots) are shown for the 
  CoRoT-7-system, together with their 1 and  $5 \sigma$ error ellipses (reddish and
greenish, 
  resp). Errors are due to the measurements {\em and} analysis of the RV signals for 
  CoRoT-7 and the high resolution spectroscopic analysis of its photosphere.
Semi-major axis
  ranges excluded by the Hill-exclusion stability criteria are shaded. Dark, reddish
and light,
  blueish areas corresponding to $k$-values of 7 and 10 respectively. Uncertainties
in stability
  analysis, orbital inclinations and measurement errors are included approximately
by the latter 
  value, see text.
  The stellar radius (yellow) and the Roche-limit (green) are indicated on the left for
  orientation. The non-overlap of the shaded areas required by
Hill-exclusion-stability 
  indicates the stability of the CoRoT-7-system as outlined here.
  Note: $1\,\sigma$ error ellipses are partially covered by the dot-symbols for
CoRoT-7b,c.}
  \label{CoRoT7HillExclusion} 
\end{figure*}

An application of this Hill-stability estimator to the CoRoT-7 system is shown in 
Fig.~\ref{CoRoT7HillExclusion}. The Hill-exclusion regions are outlined in a mass
versus 
semi-major axis diagram for the system parameters determined here. Forbidden regions
for 
other  planets in a stable system are shown by shaded areas around every orbit. The
dark reddish
shaded area is for $k=7$, a secure upper bound to the Funk et al. (2010) results. The
larger,
light-blueish areas are for $k=10$ in order to approximately account for and
securely bound 
uncertainties in the mass and semi-major axis determinations (6 and 3\% resp.\ at $5
\sigma$) 
and 15\% for a possible orbital inclination\footnote{That corresponds to an
inclination of the orbital plane 
with respect to the sky, $i_{\rm observer} = 60^\circ$.},
$i_{\rm dyn}=30^\circ$ of c and d. 
The width of the stability exclusion regions is emphasized by two horizontal bars in
the
resp.\ color at the top. The stellar radius (yellow) and the Roche-limit (green) are
plotted as 
vertical bars on the left.

For comparison the mass-estimates of the discovery paper, Q09 are 
shown with their 1 and $5 \sigma$ error-ellipses to demonstrate that they
clearly provide less stringent stability constraints due to smaller mass values.

The CoRoT-7 system as presented here, is clearly found stable as demonstrated by
the non-overlap of the appropriate Hill-exclusion zones. Not much space for stable
planet orbits
is left between components c and d. Thus the system is dynamically full --- as the
solar
system, cf. Lecar et al. (2001). However, according to this simple approach a
small planet 
might fit in between b and c if orbits are and remain circular.

This merits a detailed dynamical analysis that fully accounts
for orbital inclinations (with only loose observational constraints for c,d), the
eccentricities
and the interactions of all components on the various timescales.

\subsection{A short dynamical study of the CoRoT-7 system}

We performed a more detailed dynamical  study of the stability of  CoRoT-7 using
different initial parameters that took into account
the errors in the orbital elements:

\begin{itemize}

\item The orbital periods of all three planets have only errors in the order of 0.1
$\%$ and consequently for our numerical study these were not varied. 
\item Because of the large errors in the perihelia we just set them to 0; 
\item For the planet 
masses we have taken the nominal values $+1\sigma$ (8.3, 12.8 and 17.1 
$M_{\oplus}$ for 
CoRoT-7b, CoRoT-7c and CoRoT-7d, respectively).

\item The eccentricities of all planets are relatively small and especially the
innermost one may suffer from tides which may keep its eccentricity very 
small\footnote{The computations have been undertaken in the Newtonian framework
with assumed point masses of the planets. Internal tests have shown that the tides
may not significantly change all  orbits even when the parameter
$Q_0$ was varied within orders of magnitudes}.

\item There is no information available about the mutual inclination; therefore in a first
attempt we just set their orbital inclination to small values (1, 2 and
$3^{\circ}$). 

\end{itemize} 

With these initial conditions the orbits turned out to be very stable so  in the next step
we changed their mutual inclinations. In fact it is quite improbable 
that all three planets are moving in the same plane and the two outer planets
can move in different orbital planes with respect to CoRoT-7b. 
Two runs were tested\footnote{Note that an assumed inclination in the signal
of the RV means that the masses of CoRoT-7c and 7d are minimum masses; we
therefore in our test computations took care of this fact.}

\begin{enumerate}

\item {\bf R1}: The inclination of CoRoT-7c was initially set to $1, 16 \mbox{ and
}31^{\circ}$, whereas the other two planets initially moved in the same
  plane.
\item {\bf R2}:  The inclination of CoRoT-7d was initially set to $1, 16 \mbox{ and
}31^{\circ}$, whereas the other two planets initially moved in the same
  plane
\end{enumerate}

The results of {\bf R1} are shown in Fig.~18, where it is well visible that a
different inclination of one of the planet's orbit may change the other orbits 
drastically. For a $1^{\circ}$ inclination the three planets system is in
a very stable state, which is visible from the quasiperiodicity state of all three
orbits (Fig~18, upper panel). An additional check of the other orbital elements
show that the perihelions are moving smoothly, the nodes are in a state of
libration.  For a $16^{\circ}$ inclination the signal is already very
different but still has the sign of being quasiperiodic (regular and stable!).
This is also true for the secular motion of the nodes and perihelia. 
Even with an inclination of $31^{\circ}$ for CoRoT-7c the time evolution
of the eccentricities of all three planets is periodic and shows a signal typical
for a stable dynamical system (Fig. 18 lower panel).

\begin{figure*}
\begin{center}
\includegraphics[width=4.5in,angle=-90]{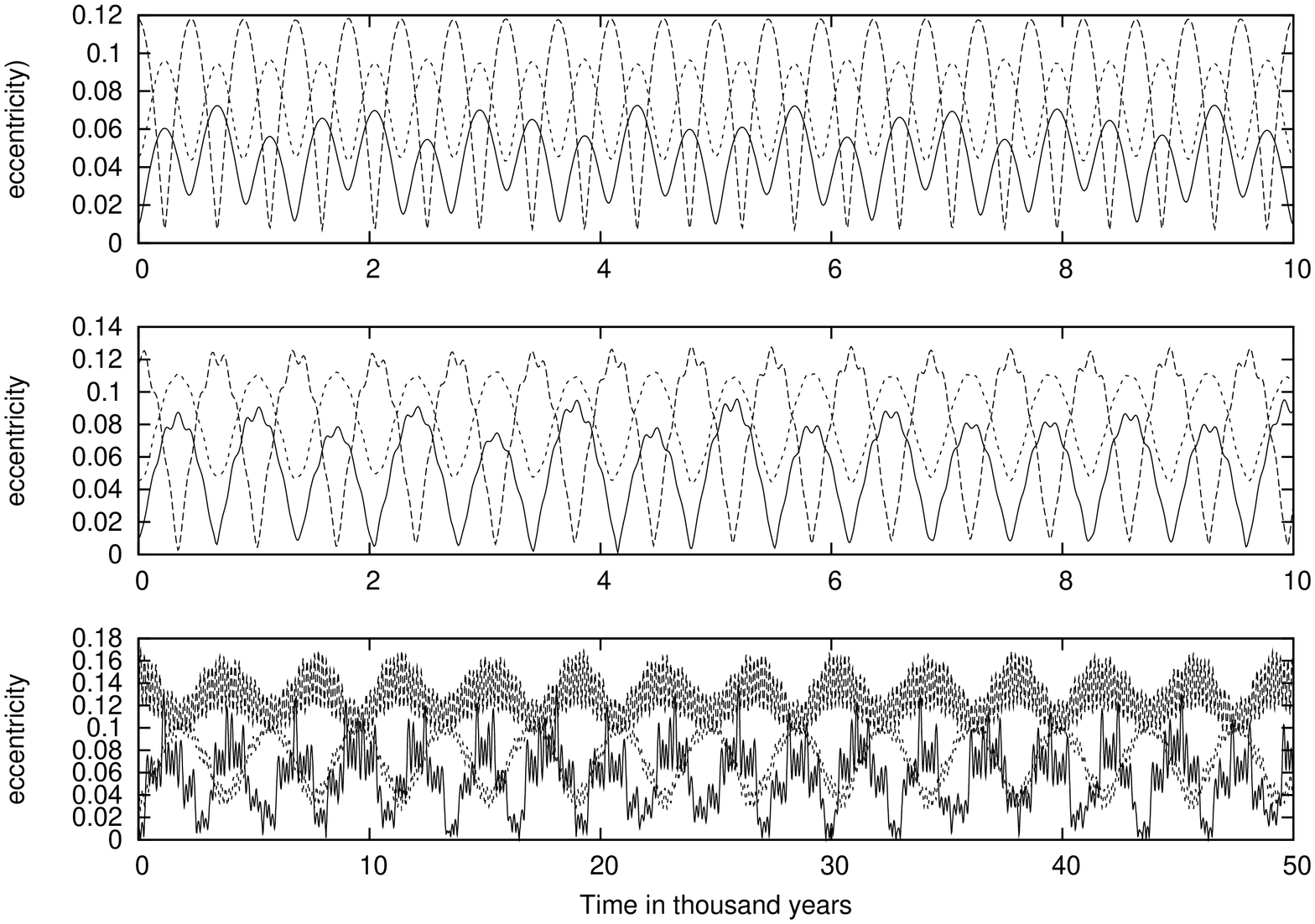}
\caption{Time evolution of the three planets' eccentricities for different
  inclinations of CoRoT-7c: 1 (top),16 (middle) and $31^{\circ}$
  (lower).}
\end{center}
\label{dyn1}
\end{figure*}

The results of {\bf R2} are shown in Fig.~19, where it is again seen that a
different inclination for one of the planets' orbits (in this case of CoRoT-7d)
may change the other orbits significantly. 
This is not the case when the inclination of the outermost planet is set to
only $1^{\circ}$; we do not show this graph since it is quite similar to the
one with CoRoT-7c set to an inclined orbit of  $1^{\circ}$.
In the two upper panels we depict the results of the time evolution of
the eccentricities of all three
planets using  an inclination for CoRoT-7d
of $16^{\circ}$  (upper panel) and $31^{\circ}$ (middle panel). In the upper
panel it is visible that the innermost planet has a quite
different period in its eccentricity behavior compared to Fig. 18 
(middle panel).
For  $i=31^{\circ}$ the picture is very different from the lower panel
in Fig. 18. It is evident that the CoRoT-7b (eccentricity $0 < e < 0.45$)
is not on a regular orbit but shows signs of chaoticity in eccentricity
and in inclination (Fig. 19 lower panel). This can be understood in terms
of the larger mass of CoRoT-7d strongly perturbing
CoRoT-7c  and thus indirectly perturbing
CoRot-7b and making the orbit of this innermost planet chaotic, but
not yet unstable. A larger inclination of either CoRoT-7c or 
CoRoT-7d may put CoRoT-7b into a
Kozai resonance (Kozai 1962) which is known to enlarge the eccentricity significantly and
can lead to unstability of the whole system.

\begin{figure*}
\begin{center}
\includegraphics[width=4.5in,angle=-90]{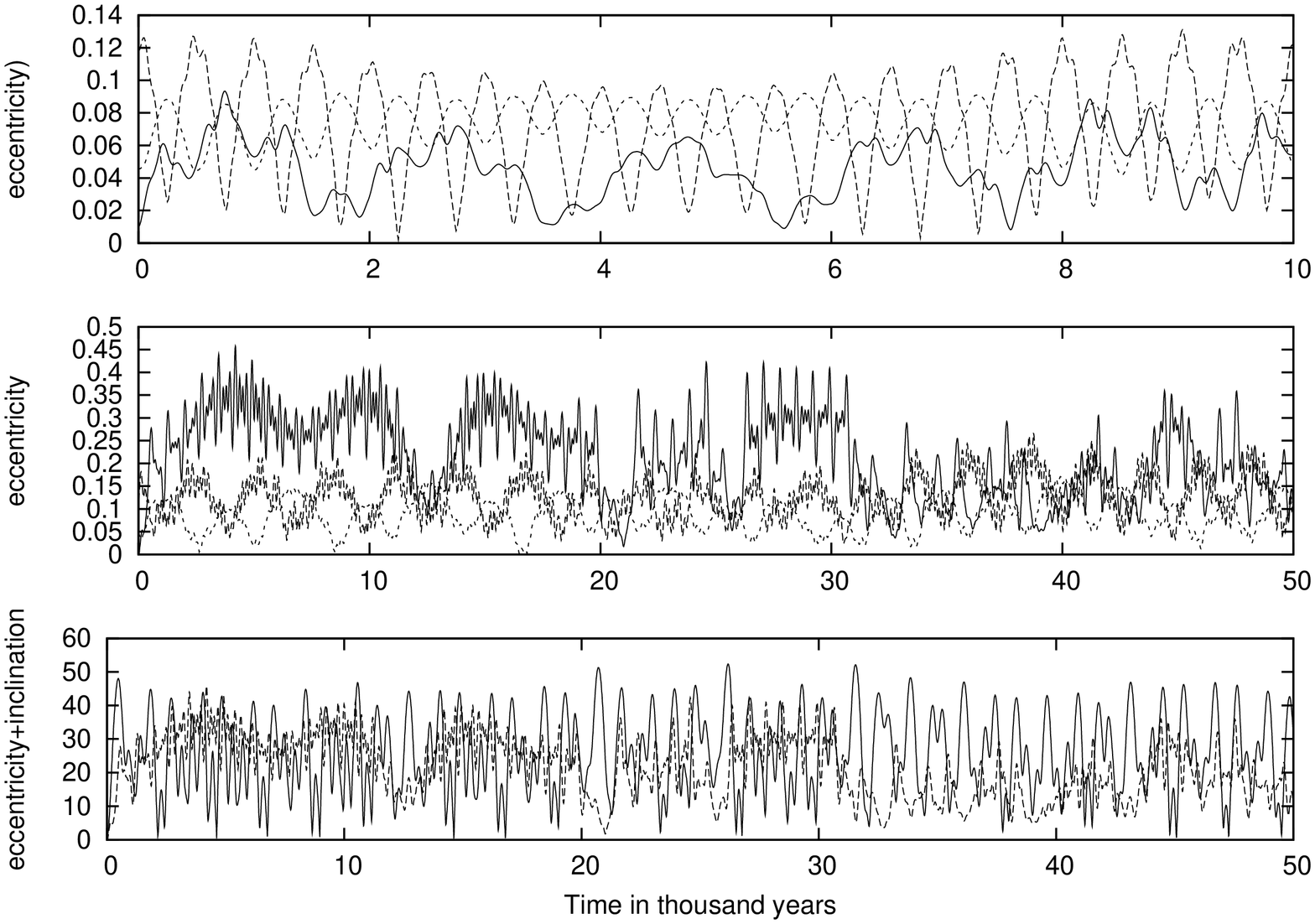}
\caption{
Time evolution of the three planets' eccentricities for different
inclinations of CoRoT-7d: $i=16^{\circ}$ (upper graph) and  $i=31^{\circ}$
(middle graph). On the lower panel we plot the inclination and the
eccentricity (times 100) of CoRoT-7b, where the nonregular behaviour is
clearly visible.}
\end{center}
\label{dyn2}
\end{figure*}

\subsubsection{Consequences of different inclinations for the transit}

With the  quite reasonable assumption that the three planets 
move on mutually inclined orbits  we now
test how that would influence the transit of the innermost planet
CoRoT-7b. We cannot expect a large
influence in its orbit on the short time scales CoRoT-7 was observed
by CoRoT or HARPS. However,
the big advantage of the planetary
system which we are studying is
that CoRoT-7b has an orbital period of less than one day 
and the other two presumed planets are also in very close-in orbits around 
the star.
This accelerates the dynamics of the system, which means that any
mutual gravitational influence (perturbations on the Keplerian orbits) 
is visible on short time scales. 
We checked the change in the inclination of the orbit of CoRoT-7b
depending on the mutual inclinations with the aim of
possibly measuring the influence on the 
inclination of CoRoT-7b which could be larger than the given errors bars 
derived  from the light curve analysis 
($i$ = 80.1$^{\circ}$ $\pm$ 0.3$^{\circ}$).

In fact, Fig.~20 (upper panel) shows
how the dynamics of the system 
 could influence the inclination of CoRoT-7b. The lower line shows the difference
 in inclination with respect to the actual assumed orbit after 1 year, the
 middle line after 3 years, and the upper line after 10 years when the
 inclination of CoRoT-7c has different inclinations between $1 \mbox{ and }
30^{\circ}$. Fig.~20 (lower panel) shows the same for a change in the
inclination of CoRoT-7d.

After these results we can conclude  the following:
after one year of observation (lowest line with '+') in none of the cases 
is the effect larger than the
errors on $i$. This is shown by the thick horizontal line which indicates
the error in the determined inclination of CoRoT-7b. The change in $i$ is
well within the measurement error.
After three years $\delta i$ could be in the
order $0.3^{\circ}$ when CoRoT-7c would be on an inclined orbit of ($i >
10^{\circ}$ with respect to CoRoT-7b). This is visible from the crossing of the
line (with stars) and the
thick line; it would be undetectable when CoRoT-7d would move on an inclined orbit.
Only after ten years of observation a significant effect 
($\delta i > 0.3^{\circ}$)
would be detectable for inclinations $i$ $>$ 3$^{\circ}$
and $i$ $>$ 13$^{\circ}$ for  runs {\bf R1} and  {\bf R2}, respectively.


We can conclude that

\begin{enumerate}

\item the system of three planets of the star CoRoT-7 with the parameters determined
via RV and the transits observed by CoRoT is in a very stable state even for 
hundreds millions of years when
their orbital planes are almost the same. They would even be stable for mutually
inclined orbits $i < 25^{\circ}$. 

\item mutually inclined orbits show quite interesting effects on
the duration of the transit of CoRoT-7b.  If  CoRoT-7c were inclined
by  $i > 10^{\circ}$ with respect to CoRoT-7b, a detectable effect could be 
observed in the change in
the inclination of CoRoT-7b via the duration of the transit  
within the lifetime of the satellite CoRoT.

\end{enumerate}

A paper concerning a detailed analysis of the influence of the inclination
shift on the duration of the transit is in preparation (Dvorak and
Schneider, in preparation).

\section{Discussion}

Our Fourier analysis of the HARPS time series of RV measurements for
CoRoT-7 reveals up to 9 frequencies. Most of these are associated with the
activity signal. However, this analysis also found 3 frequencies not associated
with activity:
$f_1$ = 1.1715 c\,d$^{-1}$ (P = 0.8535 d), $f_2$ = 0.270 c\,d$^{-1}$,
and $f_3$ = 0.1101 c\,d$^{-1}$). The first coincides with the transit 
frequency found in the  CoRoT-7 light curve  and $f_2$ coincides
with the second planet, CoRoT-7c, also reported by Q09 using the same 
data set. The final frequency $f_3$  may be due to an additional companion.

The analysis  of the RV data presented in Sections \S5 and \S6 reveals
that the CoRoT transit period  is present in the RV data with a very
high degree of statistical significance. There can be no doubt the transit-like events seen 
in the CoRoT-7 light curves are caused by a $\sim$ 2 $R_{\oplus}$ radius planet
in a 0.85-d orbit.

The mass of CoRoT-7b which is important for planet structure models depends
on the amplitude of the 0.85-d period in the  RV data. Unfortunately,
due to the relatively high activity of CoRoT-7 this amplitude depends on the
details of how the stellar signal is removed. 
In Q09 the mass of CoRoT-7b can thus be as low as
2.6 $M_{\oplus}$ and as high as 5.5 $M_{\oplus}$.

\begin{figure}
\begin{center}
\includegraphics[width=2.25in,angle=-90]{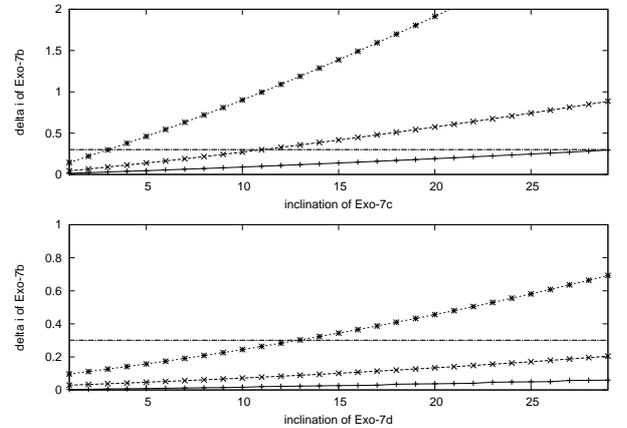}
\caption{The difference in inclination $\delta i$ of  CoRoT-7b (y-axis)
  compared to an initial plane orbit of
  CoRoT-7b (x-axis) versus the inclination of CoRoT-7c (upper graph) after 1 (lower line), 3 and 
10 years (upper line). The same plot for different inclinations of CoRoT-7d; the
thick line is the assumed error ( $\delta i = 0.30^{\circ}$ in the determination of the orbit of 
(lower graph).}
\end{center}
\label{dyn3}
\end{figure}

The analysis of the RV measurements from individual nights where at least
2 measurements were made may yield the best estimate of the RV amplitude
of CoRoT-7b. By fitting the CoRoT-7b period to the data and allowing the
zero-point velocity offset to vary each night in a least squares sense we can
account for the influence of stellar activity in a simple and model independent
way.  The fact that the calculated
zero-point offsets from each night recovers both the rotational period of
the star and the signal due to  CoRoT-7c suggests that this approach is valid. 
This approach yields an RV amplitude of 5.04 $\pm$ 1.09 m\,s$^{-1}$ which
results in a mass of 6.9 $\pm$ 1.4 $M_{\oplus}$.  

Recently, Bruntt et al. (2010) performed a detailed spectral analysis
of CoRoT-7 and determined a stellar radius of $R_{*}$ = 0.82 $\pm$ 0.04
$R_\odot$. This results in a revised planet radius of $R_p$ = 1.58 $\pm$ 0.1
$R_{\oplus}$. This along with our slightly higher planet mass results in  
a density of 9.6 $\pm$ 2.7 gm\,cm$^{-3}$ which puts it squarely 
on the Valencia et al. (2009) curve for Earth-like planets (33\% iron and
67\% silicates). In terms of structure CoRoT-7c may indeed be an Earth-like
planet.


Although this work and Q09 conclude that the 3.7-d period in the RV is due to an 
additional companion, the nature of the 9-day period is still not fully resolved. Other approaches
to the analysis of the RV time series attributes this signal to the
activity signal (Melo, in preparation).  Lanza et al. (2010) investigated
the effects of photospheric spots on the RV variations of CoRoT-7, but could
not confirm whether a 9-d period was in fact due to activity.
However, our analysis presented here gives 
strong evidence in support of a third companion:

\begin{itemize}

\item When the data is divided into subsets the Scargle power in the periodogram
at 0.11 c\,d$^{-1}$ increases with the  addition of each data set (Section \S5).
In other words the signal becomes more significant with additional data - a 
behavior  consistent with a periodic signal that is coherent and 
long-lived.

\item This signal is not found in the periodograms of any of the three activity
indicators: bisector span, Ca II S-index, and the FWHM of the CCF in spite
of all three having the exact same sampling as the RV data. Any 
activity signal found in the RV should be seen in at least one of these
indicators.

\item We have demonstrated that an analysis of subsets of the data using
filtering with rotational harmonics may remove a 9-d period to the data
even if one is known to be present. Just because harmonic filtering
removes a 9-d signal from the Fourier power spectrum is not an indication
that this signal is due to rotation modulation. 

\item The highest degree of correlation between the RV and bisector data
occurs after removing the RV signal of the 0.85-d, 3.7-d, {\it and} 9.02-d
periods. This suggests that these 3 periods do not arise from stellar activity.

\item A model of the bisector variations using scaled version of the RV
variations, but without the presence of  $f_1$, $f_2$, and $f_3$ and with
the appropriate level of noise produces a periodogram that is in good agreement
with the periodogram of the observed bisector variations. 
It is difficult to understand
why the 9-d period appears so strongly in the RV power spectrum, but not at
all in the bisector power spectrum if it is due to activity.

\end{itemize}

Our dynamical study indicates that the 3 planet system for CoRoT-7 should be stable.
More interestingly, if there are significant differences in the relative inclinations of the 
3 planets, the inclination of CoRoT-7b should change such that this could 
be measured during the mission life of CoRoT.

The hypothesis that CoRoT-7 has a third planetary companion is one that is 
easily verified. All that is required are additional RV measurements of quality
and quantity that are comparable to the HARPS measurements used in this
study. If the RV signal for CoRoT-7d remains constant in phase
and amplitude it would be difficult to attribute it to activity.
(This also holds for CoRoT-7c). However, given that the 9-d period is close to the
first harmonic of the rotational frequency, data with a rather long time base, like the
ones used in this study, may be required. Verification of the third planet is important as 
this would make the CoRoT-7 system an excellent example of an ultra-packed planetary
system.

Additional RV measurements are also important for reducing the errors on the planet mass needed
for comparing to planetary structure models.  The complexity of the 
RV variations for CoRoT-7 and the high level of activity also means that it would be difficult
to obtain a value of the planet mass of CoRoT-7b to better than about 20\% without signficantly
more RV measurements. However, CoRoT-7 is such an interesting system  further measurements
may be worthwhile. Given that 80 hours of HARPS time have already been devoted to CoRoT-7 only 
stresses the need for considerable  telescope resources 
 for precise RV studies of exoplanets.

\begin{acknowledgements}
The authors thank DLR and the German BMBF for support under grants 50~OW~0205,
50~OW~0501, 50~OW~0603. G\"unther Wuchterl thanks the MPE for computation-resources 
on DEC/VMS machines. APH would  like to thank the CoRoT Exoplanet Science Team 
for useful discussions, helpful comments, and critical remarks which resulted
in a much improved analysis of the RV data. We also thank the referee, Gordon
Walker, for useful comments which resulted in an improved paper. We also 
thank him for his rapid refereeing of the paper.
\end{acknowledgements}


\begin{thebibliography}{}

\bibitem[]{} Auvergne, M., Bodin, L., Boisnard, L. et al. 2009, A\&A, 506, 411

\bibitem[]{} Baglin, A., Auvergne, M., Boisnard, L. et al. 2006, in COSPAR, Plenary Meeting
36,  36th COSPAR Scientic Assembly, 3749

\bibitem[]{} Bruntt H., Deleuil, M., Fridlund, M., Alonso, R., Bouchy, F.,
Hatzes, A., Mayor, M., Moutou, C., \& Queloz, D. 2010, A\&A, in press.
(arXiv1005.3208v1)

\bibitem[]{Chambers} Chambers, J.~E., {Wetherill}, G.~W., \& {Boss}, A.~P. 1996, Icarus, 119, 261

\bibitem[]{} Charbonneau, D., et al. 2010, Nature, 462, 891

\bibitem[{{Funk} {et~al.}(2010){Funk}, {Wuchterl}, {Schwarz}, {Pilat-Lohinger},
  \& {Eggl}}]{Funk2010}
{Funk}, B., {Wuchterl}, G., {Schwarz}, R., {Pilat-Lohinger}, E., \& {Eggl}, S.
  2010, \aap\ {subm.}

\bibitem[]{} Hatzes, A.P., Cochran, W.J., \& Bakker, E.J. 1998, ApJ, 508, 380.

\bibitem[]{} Kozai, Y. 1962, AJ, 67.  591

\bibitem[1997]{Kur97} K\"urster, M., Schmitt, J.H.M.M., Cutispoto, G. \& Dennerl, K. 1997, A\&A, 320, 831
  
\bibitem[Kuschnig et  al.(1997)]{Kuschnig97} Kuschnig, R., Weiss, W.~W., Gruber, R., Bely, P.~Y., \& Jenkner, H.\ 1997, \aap, 328, 544 

\bibitem[]{} Lanza, A.F., Bonomo, A.S., Moutou, C., et al. A\&A, in press 
(arXiv:1005.3602v1).

\bibitem[{{Lecar} {et~al.}(2001){Lecar}, {Franklin}, {Holman}, \&
  {Murray}}]{2001ARA&A..39..581L}
{Lecar}, M., {Franklin}, F.~A., {Holman}, M.~J., \& {Murray}, N.~J. 2001,
 ARAA, 39, 581

\bibitem[]{} Leger, A., Rouan, D., Schneider, J. et al. A\&A, 506, 287.

\bibitem[]{} Lenz, P. \& Breger, M. 2004, The A-Star Puzzle, IAU Symposium
No. 224, Cambridge, UJ, Cambridge University Press, p786.


\bibitem[]{}  Queloz, D., Henry, G., Sivan, J.P., et al. 2001, A\&A, 379, 279.

\bibitem[]{}  Queloz, D., Bouchy, F., Moutou, C. et al. 2009, A\&A, 506, 303

\bibitem[]{} Saar, S.H. \& Donahue, R.A. 1997, ApJ, 485, 319

\bibitem[1982]{Sca82} Scargle, J.D. 1982, ApJ, 263, 835

\bibitem[]{} Stellingwerf, R.F. 1978, ApJ, 224, 953.

\bibitem[]{} Valencia, D., Ikoma, M, Guillot, T.,  \& Nettelmann, N. 2010,  A\&A, in press (arXiv:0907.3077v1)






\end{thebibliography}
\end{document}